\documentclass[aps,prl,twocolumn,10pt,superscriptaddress,floatfix,nofootinbib,longbibliography]{revtex4-2}
\usepackage{amsmath,amssymb,bm,graphicx,xcolor}
\usepackage[caption=false]{subfig}
\graphicspath{{panels/}}
\newcommand{\heading}[1]{\par\smallskip\noindent\textit{#1}---\ignorespaces}
\newcommand{\dd}{\mathrm{d}}

\begin{document}

\title{Entanglement Drives Common Noise into the Strong-Coupling Regime}

\author{Yizhe Zhou}
\affiliation{School of Instrumentation and Optoelectronic Engineering, Beihang University, Haidian, Beijing 100191, China}
\affiliation{Institute of Large-Scale Scientific Facility, Beihang University, Beijing 100191, China}
\author{Xusheng Lei}
\affiliation{School of Instrumentation and Optoelectronic Engineering, Beihang University, Haidian, Beijing 100191, China}
\affiliation{Institute of Large-Scale Scientific Facility, Beihang University, Beijing 100191, China}
\author{Xing Heng}
\affiliation{Hangzhou Innovation Institute, Beihang University, Hangzhou 310051, China}
\affiliation{Institute of Large-Scale Scientific Facility, Beihang University, Beijing 100191, China}
\affiliation{School of Instrumentation and Optoelectronic Engineering, Beihang University, Haidian, Beijing 100191, China}
\author{Zuoxian Wang}
\affiliation{School of Instrumentation and Optoelectronic Engineering, Beihang University, Haidian, Beijing 100191, China}
\author{Danyue Ma}
\affiliation{School of Instrumentation and Optoelectronic Engineering, Beihang University, Haidian, Beijing 100191, China}
\affiliation{Institute of Large-Scale Scientific Facility, Beihang University, Beijing 100191, China}

\begin{abstract}
Under Gaussian collective dephasing parallel to the signal, the superdecoherence of an $N$-atom Greenberger--Horne--Zeilinger (GHZ) state cancels its gain in Fisher information, so GHZ frequency sensitivity is limited to an atom-number-independent floor.
We demonstrate that this floor is a property of Gaussian diffusion: a single common phase kick can at most randomize the phase of an $N$-atom coherence, so discrete events at rate $\Gamma$ cannot dephase any coherence order faster than $2\Gamma$.
At the same single-atom coherence time, finite-rate Poisson kicks with an absolutely continuous amplitude law therefore saturate the order-$N$ decay rate and restore Heisenberg scaling.
We prove that Gaussian diffusion is the worst case for GHZ and Dicke-cat probes under this calibration, that only the Brownian component of L\'evy phase noise sets the asymptotic floor, and that the $1/N$ exponent is optimal for parallel Ramsey protocols.
These results identify the counting statistics of the common noise, rather than its spectrum, as the property that bounds superdecoherence.
\end{abstract}

\maketitle

\heading{Introduction}
Entangled probes surpass the standard quantum limit of parameter estimation~\cite{Braunstein1994,Giovannetti2006,Pezze2018}.
Spin squeezing and Greenberger--Horne--Zeilinger (GHZ) entanglement provide complementary resources~\cite{Wineland1992,Bollinger1996,Pezze2018}.
GHZ probes attain Heisenberg scaling in the absence of noise~\cite{Bollinger1996}.
Both resources are now available in ion registers, cavity-coupled ensembles, optical clocks, and programmable quantum sensors~\cite{Monz2011,Leroux2010,Eckner2023,Cao2024,Yang2025,Kaubruegger2021,Marciniak2022,Long2026}.
However, the enhanced response of a GHZ state to a frequency signal also amplifies phase noise shared by all atoms, so that its order-$N$ coherence decays $N^2$ times faster than that of a single atom.
This superdecoherence has been observed in ion registers and modeled by correlated Gaussian phase noise~\cite{Monz2011}.
Under Gaussian Markovian collective dephasing parallel to the signal, this amplification produces an atom-number-independent sensitivity floor~\cite{Andre2004,Dorner2012,KesslerClock2014,Altenburg2016,Riberi2023}.
These results assume Gaussian common noise, for which the single-atom coherence time fixes the decay of every coherence order.

Established routes beyond noise-induced precision limits exploit noise correlations or the distinguishability of signal and noise.
Temporal correlations can improve scaling through short-time Zeno dynamics~\cite{Matsuzaki2011,Chin2012,Smirne2016,Long2022,BaiAn2023}.
Quantum error correction and differential encodings protect signals distinguishable from the dominant noise~\cite{Altenburg2016,Zhou2018,Arrad2014,Kessler2014,Dur2014,Reiter2017,Zhou2024,LiuYang2025,Xu2026}, while tailored spatial correlations and nonparallel noise directions provide further routes to enhanced precision~\cite{Riberi2023,Chaves2013,Brask2015,Peng2024}.
Whether non-Gaussian common noise can remove the collective sensitivity floor while retaining Markovian dynamics and parallel signal--noise coupling has not been addressed.

Finite-rate phase kicks provide a non-Gaussian alternative to continuous phase diffusion.
Strongly coupled fluctuators, correlated bursts in solid-state arrays, and frequency jumps of reference oscillators motivate such event-based descriptions~\cite{Paladino2014,Wilen2021,McEwen2022,Harrington2025,Formichella2016}.
Compound-Poisson models retain individual phase kicks while generating Markovian dynamics~\cite{Vacchini2005,Kiely2017,Cattaneo2021,Funo2024}.
A single-atom Ramsey coherence samples the first Fourier component of the kick distribution, whereas an $N$-atom GHZ coherence samples the $N$th.
Environments with the same single-atom Ramsey coherence time can therefore have different high-order decay rates.
The persistence of the Gaussian sensitivity floor then depends on how these rates grow with atom number.

In this Letter, we compare finite-rate Poisson phase kicks with Gaussian diffusion at the same single-atom Ramsey coherence time.
For absolutely continuous kick distributions, high-order coherence decay saturates at the event rate, restoring Heisenberg scaling in GHZ Ramsey sensing.
Gaussian diffusion is the worst case for GHZ and fixed-order Dicke-cat probes.
Extending the analysis to common L\'evy phase noise shows that infinite-activity pure jumps can yield intermediate scaling and that only a Brownian component generates an asymptotic GHZ sensitivity floor.
For kick densities with finite location Fisher information, a converse establishes scaling optimality in parallel, nonadaptive Ramsey sensing.
Residual independent decoherence restricts the Heisenberg regime to a finite atom-number window.

\heading{Model and exact calibration}
We consider $N$ two-level atoms with collective spin $\bm J$ and signal Hamiltonian $H=\omega J_z$. A finite-activity common bath applies register-wide rotations $U(a)=e^{-iaJ_z}$ at Poisson rate $\Gamma$, with independent amplitudes $a$ drawn from $p(a)$. Averaging over the event record gives~\cite{Kiely2017,Funo2024}
\begin{equation}
\begin{aligned}
\dot\rho&=-i[\omega J_z,\rho]+\Gamma\bigl(\mathcal K[\rho]-\rho\bigr),\\
\mathcal K[\rho]&=\int\!\dd a\,p(a)e^{-iaJ_z}\rho e^{iaJ_z}.
\end{aligned}
\label{eq:me}
\end{equation}
The map $\mathcal K$ is a mixture of unitary channels, so Eq.~\eqref{eq:me} generates Markovian, completely positive divisible dynamics. In the $J_z$ basis, the matrix element $|m\rangle\!\langle m'|$ has coherence order $q=m-m'$. Each event multiplies it by $e^{-iaq}$, and averaging samples $\varphi(q)=\int\dd a\,p(a)e^{-iaq}$, the $q$th Fourier component of the wrapped phase-kick distribution. A single atom probes only $q=1$, a Dicke-cat superposition $(|J,q/2\rangle+|J,-q/2\rangle)/\sqrt2$ of Dicke states with $J=N/2$ probes a chosen intermediate order $q$, and the $N$-atom GHZ state probes the extremal order $q=N$.

As a concrete example used throughout, we take the symmetric Laplace law $p(a)=e^{-|a|/\mu}/(2\mu)$, for which $\varphi(q)=(1+\mu^2q^2)^{-1}$. A Gaussian bath can be matched to this one either by the single-atom Ramsey rate, which a single-atom measurement returns, or by the second-cumulant rate, which matches the two-point noise intensity, that is, the white-noise spectral density,
\begin{equation}
\begin{aligned}
\gamma_{\mathrm{s}}&\equiv\Gamma[1-\mathrm{Re}\,\varphi(1)]
=\frac{\Gamma\mu^2}{1+\mu^2}=\frac{1}{T_2},\\
D_2&\equiv\Gamma\langle a^2\rangle=2\Gamma\mu^2
=2\gamma_{\mathrm{s}}(1+\mu^2).
\end{aligned}
\label{eq:calibration}
\end{equation}
The two coincide only in the diffusion limit $\mu\to0$. Because an experiment calibrates its bath through the measured $T_2$, the Gaussian comparator relevant to metrology is matched at exact fixed $T_2$ and has $\Gamma_q^{\mathrm{G},T_2}=\gamma_{\mathrm{s}}q^2$. A Brownian comparator matched instead at fixed second cumulant has $\Gamma_q^{\mathrm{G},2}=D_2q^2/2$. We use the first convention in the Letter and label the second explicitly when it appears in the Supplemental Material (SM)~\cite{SM}.

\begin{figure}[t]
\centering
\includegraphics{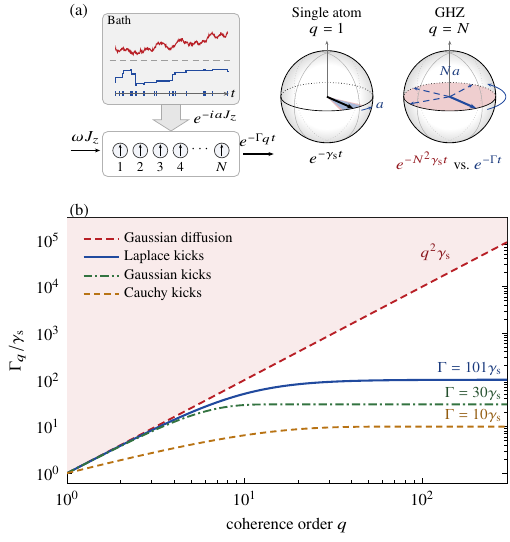}
\caption{\label{fig:mech} Mechanism at fixed single-atom $T_2$ (Laplace kicks, $\mu=0.1$). (a)~A common bath rotates the whole register by $e^{-iaJ_z}$, as Gaussian diffusion or as Poisson kicks at rate $\Gamma$, and a coherence of order $q$ decays as $e^{-\Gamma_q t}$. At $q=1$ the two models coincide. At $q=N$ and $\mu N\gg1$ a single kick rotates the GHZ phase by $Na$ and randomizes it. The surviving coherence approaches the no-kick branch $e^{-\Gamma t}$, independent of $N$, whereas diffusion gives $e^{-N^2\gamma_{\mathrm{s}}t}$. (b)~Coherence-order rate $\Gamma_q$ for three kick laws with the same $\gamma_{\mathrm{s}}$: all saturate at their event rate $\Gamma$ below the Gaussian line $q^2\gamma_{\mathrm{s}}$.}
\end{figure}

\heading{Rate saturation and GHZ metrology}
Equation~\eqref{eq:me} gives the exact coherence-decay rate
\begin{equation}
\Gamma_q=\Gamma[1-\mathrm{Re}\,\varphi(q)],\qquad 0\leq\Gamma_q\leq2\Gamma.
\label{eq:rate}
\end{equation}
Equation~\eqref{eq:rate} bounds collective decoherence at every coherence order. A single unitary event can at most randomize the phase of a coherence, so a finite mean number of events per unit time (finite activity) cannot produce a decay rate above $2\Gamma$ at any coherence order. The regime of a given probe is set by the phase it accumulates per event. A single atom accumulates the phase $a\sim\mu$ per event; for $\mu\ll1$ it decays at the weak-coupling rate $\gamma_{\mathrm{s}}\simeq\Gamma\langle a^2\rangle/2=\Gamma\mu^2$, quadratic in the phase per event. A Gaussian bath of the same $T_2$ reproduces this rate, so a single-atom measurement does not distinguish the two baths. An order-$q$ coherence accumulates the phase $qa$ per event. Beyond $q\simeq1/\mu$ it enters the strong-coupling regime of a qubit coupled to a single telegraph fluctuator or to shot noise~\cite{Paladino2002,Galperin2006,Bergli2009,NederMarquardt2007}, in which the decay rate is set by the switching or arrival rate of the noise and no longer by its strength. For any absolutely continuous kick density, the Riemann--Lebesgue lemma gives $\varphi(q)\to0$ and $\Gamma_q\to\Gamma$ as $|q|\to\infty$. Entanglement thus drives a bath that is weakly coupled to each atom into its strong-coupling regime. Whether superdecoherence persists is then decided by the crossover order $1/\mu$, which is invisible to a $T_2$ measurement. For deterministic and lattice laws $\Gamma_q$ stays bounded but may oscillate without converging; the SM gives the exception~\cite{SM}. Gaussian diffusion, by contrast, is the singular limit of infinitely many infinitesimal events, in which the crossover moves to arbitrarily large $q$ and the quadratic law holds at every finite coherence order. Figure~\ref{fig:mech}(a) contrasts the finite-activity and Gaussian baths on the single-atom and GHZ Bloch spheres, and Fig.~\ref{fig:mech}(b) shows the saturation of $\Gamma_q$ for three kick laws with the same $T_2$. The same bounded per-event action also caps collectively enhanced emission in a dissipative Dicke cascade, as shown in Sec.~S11 of the SM~\cite{Funo2024,SM}.

For the Laplace bath at fixed $T_2$, the GHZ rate is
\begin{equation}
\Gamma_N=\Gamma\frac{\mu^2N^2}{1+\mu^2N^2}
=\gamma_{\mathrm{s}}\frac{(1+\mu^2)N^2}{1+\mu^2N^2}
\xrightarrow[N\to\infty]{}\Gamma,
\label{eq:GHZrate}
\end{equation}
whereas Gaussian diffusion gives $\gamma_{\mathrm{s}}N^2$. The rates agree at $N=1$, since the two environments have the same measured Ramsey $T_2$, but separate once the probe samples higher Fourier components of the kick distribution. For $\mu N\gg1$ a single event randomizes the order-$N$ phase, so the jump bath erases the GHZ coherence outright instead of dephasing it gradually. The coherence reduces to the probability $e^{-\Gamma t}$ that no event has occurred, which does not depend on $N$; the saturation of $\Gamma_N$ at $\Gamma$ is this survival probability. Gaussian diffusion instead spreads the phase of every trajectory by an amount that grows with $N$. The same no-kick branch underlies the converse below.

\begin{figure}[t]
\centering
\includegraphics{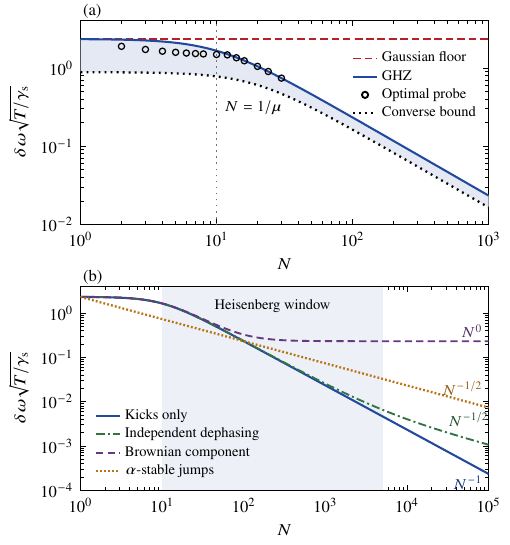}
\caption{\label{fig:metro} Optimized GHZ precision at fixed $T_2$ ($\mu=0.1$). (a)~Standard deviation $\delta\omega\sqrt{T}$ for Gaussian collective dephasing, for finite-activity kicks, and for the converse bound of Eq.~\eqref{eq:converse}; the band is the $\sqrt2$ prefactor gap. Circles: numerically optimized parallel probes for $N\leq30$ from the SM~\cite{SM}, which improve on GHZ below $N=1/\mu$ and merge with it beyond. (b)~Noise classes: pure finite activity, residual independent dephasing $\gamma_{\mathrm{i}}=0.02\gamma_{\mathrm{s}}$, a 1\% Brownian share, and an infinite-activity $\alpha$-stable phase process with $\alpha=1$. Shading marks the Heisenberg window $1/\mu\lesssim N\lesssim\Gamma/\gamma_{\mathrm{i}}$.}
\end{figure}

A GHZ Ramsey interrogation of duration $t$ has quantum Fisher information (QFI) $F_Q=N^2t^2e^{-2\Gamma_Nt}$. For $\Gamma_N>0$ and a total sensing time $T$ that permits an interior optimum, dividing $T$ into independent cycles and optimizing $F_Q/t$ gives
\begin{equation}
t^*=\frac{1}{2\Gamma_N},\qquad
\delta\omega^2T=\frac{2e\Gamma_N}{N^2}.
\label{eq:precision}
\end{equation}
The fixed-$T_2$ Gaussian model therefore produces the floor $\delta\omega^2T=2e\gamma_{\mathrm{s}}$. With finite activity, $\delta\omega\sqrt T\to\sqrt{2e\Gamma}/N$. Figure~\ref{fig:metro}(a) displays the restored standard-deviation scaling. Because the coherence decays exponentially at all times, the restored scaling requires neither bath memory nor a Zeno interval, in contrast to the non-Markovian route.

The comparison does not depend on the Laplace law. At fixed $T_2$ it holds for every kick law at every integer coherence order, because the characteristic-function inequality $1-\mathrm{Re}\,\varphi(q)\leq q^2[1-\mathrm{Re}\,\varphi(1)]$~\cite{HeathcotePitman1972} gives
\begin{equation}
\Gamma_q\leq\min\!\left(2\Gamma,q^2\gamma_{\mathrm{s}}\right).
\label{eq:extremal}
\end{equation}
For $q>1$ and $\gamma_{\mathrm{s}}>0$, the quadratic inequality is strict for every finite-activity kick law. When $\Gamma_N>0$ and $t^*\le T$, Eq.~\eqref{eq:precision} gives $\delta\omega^2T=2e\Gamma_N/N^2<2e\gamma_{\mathrm{s}}$ at $N\geq2$. The ordering also holds for a common finite sensing time, as shown in the SM~\cite{SM}. Gaussian diffusion is therefore the worst case for the GHZ or fixed-order Dicke-cat Ramsey benchmark at fixed single-atom $T_2$, attained only as the singular diffusion limit [Fig.~\ref{fig:mech}(b)]. The single-atom calibration fixes the first Fourier component of the phase law but leaves its higher components free. Diffusion attains the quadratic bound of Eq.~\eqref{eq:extremal} at every order, whereas finite events cannot amplify their per-event action without bound.

\heading{Noise-statistics classification and optimality}
The L\'evy--Khintchine form extends the result beyond compound-Poisson noise. For an arbitrary common L\'evy phase process,
\begin{equation}
\begin{aligned}
\Gamma_N&=\frac{\sigma^2N^2}{2}+\int\!\bigl(1-\cos Na\bigr)\,\dd\nu(a),\\
\frac{\Gamma_N}{N^2}&\longrightarrow\frac{\sigma^2}{2},
\end{aligned}
\label{eq:levy}
\end{equation}
where $\sigma^2$ is the Brownian coefficient and $\nu$ is the jump measure~\cite{Vacchini2005}. Finite activity gives $\Gamma_N=O(1)$ and hence GHZ $1/N$ scaling. Infinite-activity pure jumps remain subquadratic and can give intermediate scaling; for $\Gamma_N\propto N^\alpha$ with $0<\alpha<2$, $\delta\omega\propto N^{\alpha/2-1}$. A Brownian component supplies the $N^2$ term and gives the asymptotic variance floor $\delta\omega^2T\to e\sigma^2$. In terms of the Brownian share $f=\sigma^2/(2\gamma_{\mathrm{s}})$ of the single-atom rate, this floor is $2ef\gamma_{\mathrm{s}}$, the Gaussian floor reduced by the factor $f$. Jump terms can control finite-$N$ performance, but they do not generate an $N^2$-order floor.

\begin{figure}[t]
\centering
\includegraphics{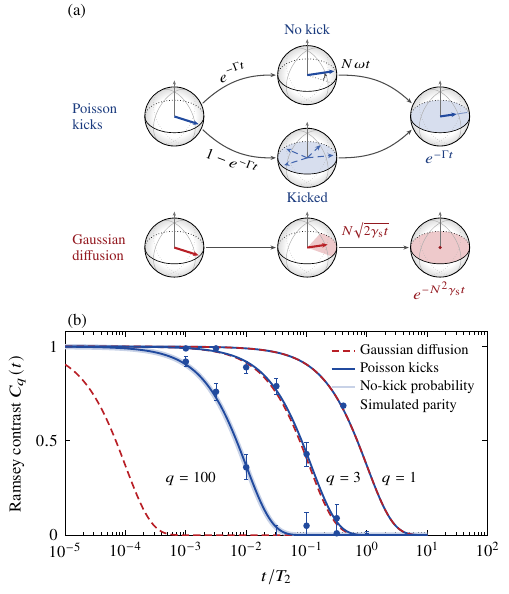}
\caption{\label{fig:origin} Physical origin and experimental signature. (a)~Branch structure of the order-$N$ coherence during one interrogation on the GHZ Bloch sphere. Under Poisson kicks the phasor splits into a no-kick branch of weight $e^{-\Gamma t}$, which stays sharp and rotates only by the signal phase $N\omega t$, and kicked branches of weight $1-e^{-\Gamma t}$, in which a single kick rotates it by $Na\gg1$ into a random direction. Their sum is a sharp phasor of length $e^{-\Gamma t}$, independent of $N$, on a randomized pedestal. Gaussian diffusion with the same $T_2$ spreads the phasor by $N\sqrt{2\gamma_{\mathrm{s}}t}$ and leaves only $e^{-N^2\gamma_{\mathrm{s}}t}$. (b)~Ramsey contrast $C_q(t)$ for $q=1$, $3$, and $100$, identical for both models at $q=1$. Under diffusion the curves shift as $q^{-2}$, whereas under finite activity they saturate at the no-kick probability $e^{-\Gamma t}$ for $q\gg1/\mu$. Points: simulated parity contrast with 200 shots per point.}
\end{figure}

The $1/N$ scaling attained by GHZ has a converse. Let $p$ be absolutely continuous with finite location Fisher information $J[p]=\int\dd a\,(p')^2/p$ ($J[p]=1/\mu^2$ for the Laplace law). Every parallel, nonadaptive Ramsey protocol with an arbitrary ancilla-assisted input, one uninterrupted interrogation, an arbitrary final positive-operator-valued measure, optimized $t$, and independent repetitions then obeys
\begin{equation}
\delta\omega^2T\geq
\left[\frac{N^2}{e\Gamma}+\frac{cJ[p]}{\Gamma}\right]^{-1},
\qquad c=1.2952\ldots .
\label{eq:converse}
\end{equation}
The proof conditions on the Poisson event count. With probability $e^{-\Gamma t}$ no kick occurs, independent of $N$, and that branch retains unitary evolution with Heisenberg-sensitive phase spread; it is the physical origin of the $N^2$ term in the available information [Fig.~\ref{fig:origin}(a)]. Kicked branches are classical location families whose information is bounded by Stam's inequality~\cite{Stam1959,SM}, and they contribute at most the $N$-independent second term. GHZ attains $\delta\omega\sqrt T\sim\sqrt{2e\Gamma}/N$, so Eq.~\eqref{eq:converse} proves asymptotic optimality of its $1/N$ scaling within the stated protocol class while leaving a factor $\sqrt2$ in the leading standard-deviation prefactor [Fig.~\ref{fig:metro}(a)]. Because the proof conditions on the event count, Eq.~\eqref{eq:converse} also bounds protocols that read a classical record of the event times after a fixed interrogation. A GHZ protocol that discards kicked cycles and interrogates for $t=1/\Gamma$ attains the $N^2$ term asymptotically, so the factor $\sqrt2$ corresponds to the information lost when the bath is not monitored. Echo sequences cannot recover this factor, because a $\pi$ pulse only reverses the sign with which later kicks enter, and a symmetric kick law is invariant under this change.

Two existing results delimit the scope of this converse. General correlated-noise bounds cover broader adaptive channel-use models~\cite{Kurdzialek2023,Kurdzialek2025}; their correlated-dephasing example uses random $\pm\epsilon$ rotations whose signs form a binary Markov chain, a structure distinct from register-wide finite-activity events. Noise that is not parallel to the signal offers a complementary route, in which the noise and signal generators no longer commute and superclassical scaling becomes possible even for Markovian Gaussian statistics~\cite{Chaves2013,Brask2015}; the present mechanism keeps the two axes parallel. These results locate the present mechanism in the higher-order statistics of parallel common noise.

\heading{Robustness and experimental diagnosis}
Independent per-atom decoherence ends the Heisenberg regime at large $N$, and a Brownian share of the common noise restores the floor. Let $\gamma_{\mathrm{i}}$ be the per-atom dephasing rate and $\gamma_{\downarrow}$ the per-atom amplitude-damping rate. The GHZ coherence decays at
\begin{equation}
\Gamma_{\mathrm{coh}}=\Gamma_N+N\left(\gamma_{\mathrm{i}}+\frac{\gamma_{\downarrow}}{2}\right).
\label{eq:robust}
\end{equation}
For independent dephasing, Eq.~\eqref{eq:precision} applies after adding $N\gamma_{\mathrm{i}}$. The interval $1/\mu\lesssim N\lesssim\Gamma/\gamma_{\mathrm{i}}$, the Heisenberg window, then exhibits the finite-activity $1/N$ regime before crossing to a standard-quantum-limit (SQL) tail. Amplitude damping has the same scaling crossover, but leaked populations modify the QFI prefactor; the exact block-resolved expression is given in the SM~\cite{SM}. A Brownian share contributes the $\sigma^2N^2/2$ term in Eq.~\eqref{eq:levy} and restores an atom-number-independent floor. Figure~\ref{fig:metro}(b) separates these two limits and places the intermediate scaling of infinite-activity jumps between them.

Entangled atomic clocks are a natural place to look for this window, because the clock signal and the common local-oscillator (LO) phase noise are both proportional to $J_z$~\cite{Andre2004,Borregaard2013,KesslerClock2014,Komar2014,Ludlow2015,Schulte2020,Nichol2022}. For a jump-dominated LO with event rate $\Gamma$, kick scale $\mu$, and residual independent rate $\gamma_{\mathrm{i}}$, the size of the window depends on the Brownian share and on preparation, readout, and dead-time costs. The SM gives an idealized parameter example over this window~\cite{SM}. Event-like common environments are also motivated by strongly coupled fluctuators and correlated quasiparticle or radiation bursts in solid-state arrays~\cite{Paladino2014,Wilen2021,McEwen2022,Harrington2025}. Classical diagnostics of such records are collected in the SM~\cite{SM}.

The saturation of $\Gamma_q$ with coherence order can be tested directly, without inferring a quantum response from an external noise record. Programmable register-wide phase rotations, or Poisson-timed pulses of a common magnetic field, can realize a chosen event rate and amplitude distribution. Preparing GHZ or Dicke-cat coherences of several orders then gives $\Gamma_q$ from their Ramsey contrast, so quadratic growth can be distinguished from finite-activity saturation [Fig.~\ref{fig:origin}(b)]. The quadratic growth of $\Gamma_q$ under Gaussian field noise has been measured with GHZ states of up to eight ions~\cite{Monz2011}, and the dependence of decoherence on coherence order has been measured in nuclear magnetic resonance~\cite{KrojanskiSuter2004}; the finite-activity signature is the departure from $q^2$ near $q\simeq1/\mu$. For the symmetric kick laws considered here, the measured contrast directly samples $\operatorname{Re}\varphi(q)=\varphi(q)$. The same apparatus can verify the predicted precision crossover using parity or another calibrated collective readout.

\heading{Conclusion}
The collective Gaussian floor originates from the diffusion statistics of common phase noise. At fixed single-atom $T_2$, a finite event rate caps the decay of every coherence order at $2\Gamma$; GHZ Ramsey sensing then recovers $1/N$ scaling, and every order $q\geq2$ dephases more slowly than under Gaussian diffusion. For general L\'evy noise, only a Brownian component produces an asymptotic floor, while infinite-activity jumps can give intermediate exponents. The no-kick branch underlies a converse that establishes the optimality of the Heisenberg exponent for parallel, nonadaptive Ramsey protocols, and independent noise sets the end of the Heisenberg window. Measuring $\Gamma_q$ as a function of coherence order therefore makes the higher-order statistics of common noise an accessible design parameter for collective quantum sensors.

\begin{acknowledgments}
\heading{Acknowledgments}This work is supported by the National Natural Science Foundation of China under Grant No.~42388101.
\end{acknowledgments}

\heading{Data availability}The classical noise diagnostics in the SM~\cite{SM} use publicly archived records: the fiber-network comparison of ultrastable lasers of Ref.~\cite{Schioppo2022}, LIGO auxiliary-channel data released through the Gravitational Wave Open Science Center~\cite{GWOSC2021}, and the published superconducting-array records of Refs.~\cite{McEwen2022,Li2025}. The analysis code and derived records are available from the authors.

\bibliography{References}

\end{document}


\title{Supplemental Material:\\ Entanglement Drives Common Noise into the Strong-Coupling Regime}
\author{Yizhe Zhou}
\affiliation{School of Instrumentation and Optoelectronic Engineering, Beihang University, Haidian, Beijing 100191, China}
\affiliation{Institute of Large-Scale Scientific Facility, Beihang University, Beijing 100191, China}
\author{Xusheng Lei}
\affiliation{School of Instrumentation and Optoelectronic Engineering, Beihang University, Haidian, Beijing 100191, China}
\affiliation{Institute of Large-Scale Scientific Facility, Beihang University, Beijing 100191, China}
\author{Xing Heng}
\affiliation{Hangzhou Innovation Institute, Beihang University, Hangzhou 310051, China}
\affiliation{Institute of Large-Scale Scientific Facility, Beihang University, Beijing 100191, China}
\affiliation{School of Instrumentation and Optoelectronic Engineering, Beihang University, Haidian, Beijing 100191, China}
\author{Zuoxian Wang}
\affiliation{School of Instrumentation and Optoelectronic Engineering, Beihang University, Haidian, Beijing 100191, China}
\author{Danyue Ma}
\affiliation{School of Instrumentation and Optoelectronic Engineering, Beihang University, Haidian, Beijing 100191, China}
\affiliation{Institute of Large-Scale Scientific Facility, Beihang University, Beijing 100191, China}
\maketitle

\tableofcontents
\clearpage

\section{Poisson-kick master equation and the Gaussian limit}
\label{sec:Smodel}
We model a common white non-Gaussian bath as a classical noise $\xi(t)$ coupled through the collective operator $A=A^\dagger$, $H_{\rm tot}(t)=H+A\,\xi(t)$. Here $\xi(t)=\sum_i a_i\,\delta(t-t_i)$ is a compound-Poisson (shot-noise) process, whose jump times $\{t_i\}$ are Poisson distributed with rate $\Gamma$ and whose amplitudes $\{a_i\}$ are i.i.d.\ with density $p(a)$. The conditional propagator is a free evolution under $H$ punctuated by instantaneous kicks $U(a_i)=e^{-ia_iA}$. Averaging the stochastic map over the noise on an interval $\dd t$, a kick occurs with probability $\Gamma\,\dd t$ (amplitude $a\sim p$) and is absent with probability $1-\Gamma\,\dd t$, giving
\begin{equation}
\rho(t+\dd t)=(1-\Gamma\,\dd t)\,e^{-iH\dd t}\rho\, e^{iH\dd t}+\Gamma\,\dd t\!\int\!\dd a\,p(a)\,e^{-iaA}\rho\,e^{+iaA}+O(\dd t^2),
\end{equation}
hence the time-local master equation of the main text,
\begin{equation}
\dot\rho=-i[H,\rho]+\Gamma\big(\mathcal{K}[\rho]-\rho\big),\qquad
\mathcal{K}[\rho]=\int_{-\infty}^{\infty}\!\dd a\,p(a)\,e^{-iaA}\rho\,e^{+iaA}.
\label{eq:Sme}
\end{equation}
Because $\mathcal{K}$ is a convex mixture of unitary conjugations, it is completely positive and trace preserving, and $\Gamma(\mathcal{K}-\mathbb{I})$ is a Gorini--Kossakowski--Sudarshan--Lindblad generator. The resulting semigroup is CP-divisible (Markovian). Individual coherence amplitudes have a single exponential decay in the commuting pure-dephasing case of Sec.~S2; this does not follow for arbitrary $H$ and $A$ from the semigroup property alone. In the eigenbasis of $A$, the kick map is determined by the characteristic function
\begin{equation}
\varphi(q)=\int_{-\infty}^{\infty}\!\dd a\,p(a)\,e^{-iaq},\qquad \varphi(q)=\frac{1}{1+\mu^2q^2}\ \ \text{for the symmetric Laplace law}\ \ p(a)=\tfrac{1}{2\mu}e^{-|a|/\mu}.
\label{eq:Sphi}
\end{equation}

\emph{Gaussian limit.} Expanding the kick to second order, $e^{-iaA}\rho\,e^{iaA}=\rho-ia[A,\rho]-\tfrac{a^2}{2}[A,[A,\rho]]+O(a^3)$, and using $\langle a\rangle=0$, $\langle a^2\rangle=2\mu^2$ for the symmetric law,
\begin{equation}
\Gamma(\mathcal{K}-\mathbb{I})\rho=2\Gamma\mu^2\,\mathcal{D}[A]\rho\;+\;O(\Gamma\mu^4),\qquad \mathcal{D}[A]\rho=A\rho A-\tfrac12\{A^2,\rho\},
\end{equation}
where we used $\mathcal{D}[A]\rho=-\tfrac12[A,[A,\rho]]$ for Hermitian $A$. Taking $\mu\to0$, $\Gamma\to\infty$ at fixed second-cumulant rate $D_2\equiv\Gamma\langle a^2\rangle=2\Gamma\mu^2$ yields standard Gaussian collective dephasing $D_2\mathcal{D}[A]$. The displayed remainder is $O(\mu^2)$ at fixed finite-dimensional $A$. For an order-$q$ coherence the correction is $O(\Gamma\mu^4q^4)$ and the expansion requires $\mu|q|\ll1$, so the fixed-order diffusion limit is not uniform in coherence order.

For an asymmetric law with an existing first moment, the first-order term is the Hamiltonian drift $-i\Gamma\langle a\rangle[A,\rho]$. Away from the small-kick limit, the full imaginary part of the characteristic function produces an order-dependent phase [Eq.~\eqref{eq:Scoherence}], not in general a single frequency renormalization. For a two-branch probe, a known extra phase shifts the working point without changing the single-parameter QFI. For a multi-order probe, the full complex $\varphi(q)$ must be retained. An uncalibrated phase can bias the estimator; the symmetric Laplace law avoids this issue.

\emph{Calibration conventions.} Matching a Gaussian bath to a non-Gaussian bath by its two-point noise intensity and matching it by its exact single-probe Ramsey time are distinct operations. For the symmetric Laplace compound-Poisson process,
\begin{equation}
 \gamma_{\mathrm{s}}\equiv\Gamma[1-\varphi(1)]
 =\frac{\Gamma\mu^2}{1+\mu^2}=\frac{1}{T_2},
 \qquad
 D_2\equiv\Gamma\langle a^2\rangle=2\Gamma\mu^2
 =2\gamma_{\mathrm{s}}(1+\mu^2).
 \label{eq:Scalibration}
\end{equation}
The Gaussian comparator at exact fixed $T_2$ has $\Gamma_q^{\mathrm{G},T_2}=\gamma_{\mathrm{s}}q^2$ and is generated by $2\gamma_{\mathrm{s}}\mathcal{D}[A]$. Since $\xi(t)$ is white with spectral density $D_2$ and the fixed-$T_2$ comparator has spectral density $2\gamma_{\mathrm{s}}$, the spectra of the two baths differ by the factor $1+\mu^2$. The fixed-second-cumulant Brownian comparator has $\Gamma_q^{\mathrm{G},2}=D_2q^2/2$ and is generated by $D_2\mathcal{D}[A]$. The two Gaussian calibrations coincide only in the diffusion limit, when $\gamma_{\mathrm{s}}\to D_2/2$. A nontrivial finite kick law with finite second moment cannot match both calibrations exactly because $1-\cos a\le a^2/2$, with equality only at $a=0$. The main metrological comparison uses exact fixed $T_2$; any fixed-second-cumulant comparison is labeled explicitly and requires a finite second moment.

\section{Common dephasing is diagonal: the saturation theorem}
\label{sec:Ssaturation}
For $A=J_z$ both $-i[H,\cdot]$ (with $H=\omega J_z$) and $\mathcal{K}$ are diagonal in the eigenbasis $\{|m\rangle\}$ of $J_z$, since $e^{-iaJ_z}\rho\,e^{iaJ_z}$ has matrix elements $e^{-ia(m-m')}\rho_{mm'}$. Equation~\eqref{eq:Sme} therefore decouples,
\begin{equation}
\dot\rho_{mm'}=\big[-i\omega q+\Gamma(\varphi(q)-1)\big]\rho_{mm'},\qquad q=m-m',
\end{equation}
and hence
\begin{equation}
\rho_{mm'}(t)=\rho_{mm'}(0)\,e^{-i\omega q t}\,e^{-\Gamma_q t}\,
e^{+i\Gamma t\,\mathrm{Im}\,\varphi(q)},
\label{eq:Scoherence}
\end{equation}
with an exponential decay of its magnitude at rate
\begin{equation}
\Gamma_q=\Gamma\,[1-\mathrm{Re}\,\varphi(q)],\qquad 0\le\Gamma_q\le2\Gamma.
\label{eq:SGk}
\end{equation}
The bound follows from $|\varphi|\le1$. For an absolutely continuous ($L^1$) kick density the Riemann--Lebesgue lemma gives $\varphi(q)\to0$ as $|q|\to\infty$, hence
\begin{equation}
\Gamma_q\xrightarrow[\;|q|\to\infty\;]{}\Gamma
\end{equation}
for \emph{any absolutely continuous} kick law, so the collective dephasing rate saturates at the event rate $\Gamma$. Only the small-$q$ behavior and the crossover scale depend on the distribution (Table~\ref{tab:phi}). In the diffusion limit $\mu\to0$ of this family the crossover $q^\ast\sim1/\mu$ moves to infinity and the fixed-second-cumulant rate $\Gamma_q^{\mathrm{G},2}=D_2q^2/2$ never saturates.

\begin{table}[ht]
\caption{\label{tab:phi} Representative absolutely continuous kick-amplitude laws within the compound-Poisson model: all saturate, with the crossover order $q^\ast$ (where $\Gamma_q$ reaches $\sim\!\Gamma/2$) set by the law. No common fixed-second-cumulant or fixed-$T_2$ calibration is implied by the parameter $\mu$; Eq.~\eqref{eq:Scalibration} gives both conventions for the Laplace example.}
\begin{ruledtabular}
\begin{tabular}{lccc}
Kick law $p(a)$ & $\mathrm{Re}\,\varphi(q)$ & Small-$q$ rate $\Gamma_q$ & Crossover $q^\ast$\\
\hline
Laplace $\tfrac{1}{2\mu}e^{-|a|/\mu}$ & $1/(1+\mu^2q^2)$ & $\Gamma\mu^2q^2$ & $1/\mu$\\
Gaussian (var.\ $\mu^2$) & $e^{-\mu^2q^2/2}$ & $\tfrac12\Gamma\mu^2q^2$ & $1/\mu$\\
sym.\ $\alpha$-stable (L\'evy) & $e^{-|\mu q|^\alpha}$ & $\Gamma|\mu q|^\alpha$ & $1/\mu$\\
\end{tabular}
\end{ruledtabular}
\end{table}

The last row describes a \emph{finite-activity compound-Poisson process} whose individual marks have an $\alpha$-stable density; its exponent $\Gamma[1-e^{-|\mu q|^\alpha}]$ saturates. This differs from taking the accumulated phase itself to be an $\alpha$-stable L\'evy process, which has infinite activity and an exponent proportional to $|q|^\alpha$ that does not saturate (Sec.~S4).

For a GHZ probe the relevant coherence order is $q=N$. In the main-text notation, Laplace kicks give
$\Gamma_N=\Gamma\mu^2N^2/(1+\mu^2N^2)
=\gamma_{\mathrm{s}}(1+\mu^2)N^2/(1+\mu^2N^2)$ [Eq.~(4)].
At fixed $t$, its contrast satisfies
\begin{equation}
 C_N(t)=e^{-\Gamma_Nt}
 =e^{-\Gamma t}\exp\!\left(\frac{\Gamma t}{1+\mu^2N^2}\right)
 \longrightarrow e^{-\Gamma t}.
 \label{eq:Svoidcorrection}
\end{equation}
The limiting contrast is the no-kick probability. At finite $N$, kicked trajectories still contribute.

\emph{Sufficient condition and discrete exceptions.} Absolute continuity is sufficient for the limit $\Gamma_q\to\Gamma$. More generally, the same conclusion holds whenever $\varphi(q)\to0$ along the relevant coherence orders (the Rajchman condition used in Sec.~S10). For deterministic kicks $p(a)=\delta(a-a_0)$, $\Gamma_q=\Gamma[1-\cos(a_0q)]$ remains bounded but need not converge. If $a_0=2\pi$, every integer order has $\Gamma_q=0$. Other lattice laws can give periodic or nonconvergent rates and protected subsequences. The bound $0\le\Gamma_q\le2\Gamma$ holds in all these cases; the limit equal to $\Gamma$ requires the stated additional condition.

\section{GHZ quantum Fisher information and Ramsey optimization}
\label{sec:Sramsey}
Under collective dephasing, the GHZ state $(|J,J\rangle+|J,-J\rangle)/\sqrt2$ occupies only $m=\pm J$, with $J=N/2$. Its populations remain $1/2$, and its off-diagonal element is given by Eq.~\eqref{eq:Scoherence} at $q=N$. A known asymmetric-kick phase can be included in the working point. For the symmetric case, the effective qubit has Bloch vector $\bm r=v(\cos\phi,\sin\phi,0)$, with $v=e^{-\Gamma_Nt}$ and $\phi=N\omega t$. Since $\bm r\cdot\partial_\omega\bm r=0$, the qubit formula $F_Q=|\partial_\omega\bm r|^2+(\bm r\cdot\partial_\omega\bm r)^2/(1-|\bm r|^2)$ gives
\begin{equation}
 F_Q(t)=N^2t^2e^{-2\Gamma_Nt}.
 \label{eq:SFQ}
\end{equation}

\emph{Measurement and local estimation.} A parity measurement with a controllable analysis phase $\phi_{\rm a}$ has probabilities $P_\pm=[1\pm v\cos(N\omega t+\phi_{\rm a})]/2$. At a quadrature working point its classical Fisher information equals Eq.~\eqref{eq:SFQ}. The frequency likelihood is periodic with period $2\pi/(Nt)$, so this local precision presumes prior localization or a preliminary estimate that identifies the relevant fringe. The preparation, readout, and localization costs are outside the sensing time considered here.

\emph{Optimization with a finite sensing time.} For independent cycles of duration $t$, the quantum Cram\'er--Rao bound is $\delta\omega^2\ge t/[TF_Q(t)]$. As in the main text, we neglect overhead and treat $T/t$ as effectively continuous; the resulting information bound is asymptotically attainable in the many-cycle limit. Define the optimized variance-time benchmark
\begin{equation}
 \mathcal V_N(T)\equiv
 \left[\max_{0<t\le T}\frac{F_Q(t)}{t}\right]^{-1}
 =\frac{e^{2\Gamma_Nt_{\rm opt}}}{N^2t_{\rm opt}},
 \qquad
 t_{\rm opt}=
 \begin{cases}
 \min\{T,(2\Gamma_N)^{-1}\},&\Gamma_N>0,\\
 T,&\Gamma_N=0.
 \end{cases}
 \label{eq:SfiniteT}
\end{equation}
When $\Gamma_N>0$ and $T\ge(2\Gamma_N)^{-1}$, this reduces to the main-text benchmark
\begin{equation}
 t^*=\frac1{2\Gamma_N},\qquad
 \delta\omega^2T=\frac{2e\Gamma_N}{N^2}.
 \label{eq:Sdw}
\end{equation}
For a protected order with $\Gamma_N=0$, Eq.~\eqref{eq:SfiniteT} gives $\mathcal V_N(T)=1/(N^2T)$; Eq.~\eqref{eq:Sdw} does not apply there.

At exact fixed $T_2$, Gaussian common dephasing has $\Gamma_N^{\mathrm{G},T_2}=\gamma_{\mathrm{s}}N^2$ and hence the long-time standard-deviation floor $\sqrt{2e\gamma_{\mathrm{s}}}$. For an absolutely continuous finite-activity kick law, $\Gamma_N\to\Gamma$ and the corresponding value approaches $\sqrt{2e\Gamma}/N$. The two models agree at order one. The fixed-second-cumulant comparison uses $D_2N^2/2$ and matches the small-kick expansion to leading order.

\emph{Finite-order correction to the no-kick picture.} Main-text Fig.~3(a) depicts the order-$N$ coherence as a narrow no-kick branch of weight $e^{-\Gamma t}$ on a randomized pedestal. By Eq.~\eqref{eq:Svoidcorrection}, the kicked trajectories contribute $C_N(t)-e^{-\Gamma t}=e^{-\Gamma t}\,[e^{\Gamma t/(1+\mu^2N^2)}-1]$ to the coherence factor; this contribution is positive at every finite order and vanishes only as $\mu N\to\infty$. Correspondingly, for Laplace kicks $t^*=[1+(\mu N)^{-2}]/(2\Gamma)$, so the saturated value $t_\infty=1/(2\Gamma)$ is asymptotically optimal. At $N=50$, $\mu=0.1$, the exact optimum is $1.04t_\infty$ and $C_N(t_\infty)=0.6183$, compared with the no-kick weight $e^{-1/2}=0.6065$, a relative correction of $1.94\%$ at $\mu N=5$. At the largest order shown in Fig.~3(b), $q=100$, the correction at the same $\Gamma t=1/2$ is $0.50\%$.

A coherent spin state occupies more than two Dicke levels. Its QFI must be obtained from the full output matrix, as in Secs.~S5 and S13; it cannot be assigned a single coherence order. The constructive GHZ result above is separated from arbitrary-probe optimality in Sec.~S5 and from the exact Dicke-cat family comparison in Sec.~S7.

\section{Coherence-order extremality of the Gaussian limit}
\label{sec:Sextremality}
\emph{Theorem (coherence-order bound).} For any kick law $p$, symmetric or not, and any integer $q\ge1$,
\begin{equation}
1-\mathrm{Re}\,\varphi(q)\ \le\ q^2\,\big[1-\mathrm{Re}\,\varphi(1)\big],
\qquad\text{i.e.}\qquad
\Gamma_q\ \le\ q^2\,\gamma_{\mathrm{s}},\qquad \gamma_{\mathrm{s}}\equiv\Gamma[1-\mathrm{Re}\,\varphi(1)].
\label{eq:Sextremal}
\end{equation}
\emph{Proof.} $|\sin q\theta|\le q\,|\sin\theta|$ for all real $\theta$ and integer $q\ge1$, by induction on $q$: $|\sin(q{+}1)\theta|=|\sin q\theta\cos\theta+\cos q\theta\sin\theta|\le|\sin q\theta|+|\sin\theta|$. Hence, pointwise in the kick amplitude $a$,
\begin{equation}
1-\cos qa\;=\;2\sin^2\!\frac{qa}{2}\;\le\;2q^2\sin^2\!\frac{a}{2}\;=\;q^2\,(1-\cos a).
\end{equation}
Since $\mathrm{Re}\,\varphi(q)=\mathbb{E}[\cos qa]$ for any (also asymmetric) law, averaging over $p$ gives Eq.~\eqref{eq:Sextremal}. $\square$

The inequality is classical in probability theory; it appears, in the sharper form $1-u(qt)\le q[1-u^q(t)]\le q^2[1-u(t)]$ for the real part $u$ of any characteristic function, in Ref.~\cite{S_HeathcotePitman}. Combined with the bound $\Gamma_q\le2\Gamma$ of Sec.~S2 it gives Eq.~(6) of the main text, $\Gamma_q\le\min(2\Gamma,\,q^2\gamma_{\mathrm{s}})$.

\emph{Equality analysis.} Pointwise equality at some $q\ge2$ requires $a\in2\pi\mathbb{Z}$; a law supported there has $\cos a=1$ almost surely and hence $\gamma_{\mathrm{s}}=0$. Therefore for every law with $\gamma_{\mathrm{s}}>0$ the inequality is \emph{strict} at every $q\ge2$: its order-$q$ coherence rate lies strictly below $q^2\gamma_{\mathrm{s}}$. The bound is attained in the diffusion limit, since along the Laplace family at fixed $\gamma_{\mathrm{s}}$ ($\mu\to0$, $\Gamma\to\infty$), $\Gamma_q/(q^2\gamma_{\mathrm{s}})\to1$ for every fixed $q$; the supremum is the limiting diffusion process itself, which no finite-rate member of the class attains. No positive universal lower bound exists, since the lattice law $p(a)=\delta(a-2\pi/N)$ has $\Gamma_N=0$ while $\gamma_{\mathrm{s}}=\Gamma[1-\cos(2\pi/N)]>0$; the order-$N$ coherence is decoherence-free while single atoms dephase.

\emph{GHZ-Ramsey consequence, including protected orders.} At every common interrogation time $t>0$, the inequality $\Gamma_N\le\gamma_{\mathrm{s}}N^2$ gives $F_Q^{\rm kick}(t)\ge F_Q^{\rm G}(t)$. Maximizing each information rate over the same interval $0<t\le T$ therefore gives $\mathcal V_N^{\rm kick}(T)\le\mathcal V_N^{\rm G}(T)$, with strict improvement for $N\ge2$ and $\gamma_{\mathrm{s}}>0$. When the optimum is interior and $\Gamma_N>0$, Eq.~\eqref{eq:Sdw} yields $2e\Gamma_N/N^2<2e\gamma_{\mathrm{s}}$. For a protected order, Eq.~\eqref{eq:SfiniteT} gives $1/(N^2T)$. Gaussian diffusion is the worst case for the \emph{GHZ or fixed-order Dicke-cat benchmark}; at any fixed order the supremum is approached in the diffusion limit. Optimization over arbitrary inputs is the distinct problem treated in Sec.~S5.

\begin{figure}[ht]
\centering
\subfloat[]{\includegraphics{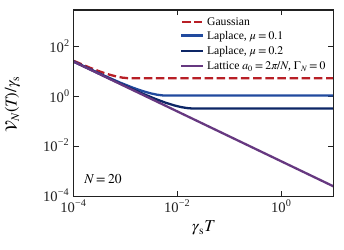}}\hspace{3pt}%
\subfloat[]{\includegraphics{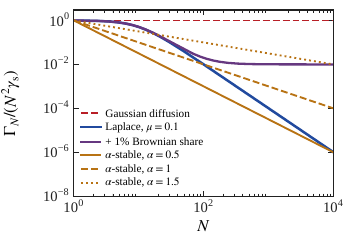}}\hspace{3pt}%
\subfloat[]{\includegraphics{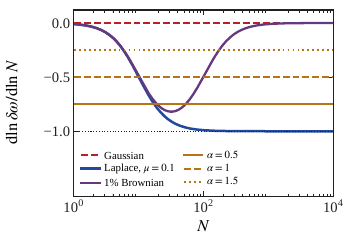}}
\caption{\label{fig:Sbenchmark} Finite-time benchmark and L\'evy classes at fixed $\gamma_{\mathrm{s}}$. (a)~Variance--time benchmark $\mathcal V_N(T)$ of Eq.~\eqref{eq:SfiniteT} for $N=20$: Gaussian dephasing, Laplace kicks with $\mu=0.1$ and $0.2$, and the protected lattice law $a_0=2\pi/N$ ($\Gamma_N=0$). All follow $1/(N^2T)$ at short $T$; the kick curves stay below the Gaussian curve at every $T$, and the protected order never leaves this branch. (b)~Normalized rate $\Gamma_N/(N^2\gamma_{\mathrm{s}})$ for Laplace kicks ($\mu=0.1$), the same jumps with a $1\%$ Brownian share, $\alpha$-stable phase processes with $\alpha=0.5$, $1$, $1.5$, and Gaussian diffusion. (c)~Local scaling exponent $\mathrm{d}\ln\delta\omega/\mathrm{d}\ln N$ of the optimized GHZ precision for the same models: $-1$ for finite activity, $\alpha/2-1$ for stable processes, $0$ on a floor.}
\end{figure}

\emph{L\'evy form and the GHZ asymptotic floor.} Let the accumulated common phase $\Theta_t$ be an arbitrary L\'evy process with triplet $(b,\sigma^2,\nu)$, $\int\min(1,a^2)\,\dd\nu(a)<\infty$. Then $\mathbb{E}[e^{-iq\Theta_t}]=e^{-t\psi(q)}$ with
\begin{equation}
\Gamma_q=\mathrm{Re}\,\psi(q)=\tfrac12\sigma^2q^2+\int\big(1-\cos qa\big)\,\dd\nu(a),
\end{equation}
the drift $b$ and the odd part of $\nu$ contributing a generally order-dependent phase. For GHZ or a fixed-order Dicke cat, a known phase shifts the working point and does not change the single-parameter QFI; for a multi-order input it must be retained in the full output state and can affect the QFI. In all cases it must be calibrated to avoid bias. The compound-Poisson bath of the main text is the finite-activity case $\nu=\Gamma\,p$; this identification of decoherence rates with L\'evy exponents is the coherence-order analogue of the position-space theory of decoherence by scattering events~\cite{S_GallisFleming,S_HornbergerSipe,S_Vacchini}. Using $1-\cos qa\le\min(2,\,q^2a^2/2)$ and dominated convergence,
\begin{equation}
\frac{1}{q^2}\int\big(1-\cos qa\big)\,\dd\nu(a)\ \le\ \int\min\!\Big(\frac{2}{q^2},\,\frac{a^2}{2}\Big)\dd\nu(a)\ \xrightarrow[\;q\to\infty\;]{}\ 0,
\end{equation}
which holds for \emph{any} L\'evy measure, of finite or infinite activity, because for $q\ge1$ the integrand is dominated by the integrable function $\min(2,a^2/2)$ and converges pointwise to zero. Hence $\Gamma_N/N^2\to\sigma^2/2$ and the optimized GHZ protocol obeys
\begin{equation}
\lim_{N\to\infty}\ \delta\omega^2\,T\ =\ e\,\sigma^2.
\label{eq:Slevyfloor}
\end{equation}
The same limit holds for any fixed $T>0$ with the constrained benchmark in Eq.~\eqref{eq:SfiniteT}. On a time-capped subsequence, $\mathcal V_N(T)\le e/(N^2T)\to0$; such a subsequence can persist asymptotically only when $\sigma^2=0$.
As in main-text Eq.~(7), the asymptotic \emph{GHZ} sensitivity floor is set by the diffusive (Brownian) component of the common noise alone. For an arbitrary parallel probe, the Brownian component alone gives the input-independent bound $\delta\omega^2T\ge\sigma^2$ by the argument of Sec.~S9 [Eq.~\eqref{eq:Sgaussianparallel} with $2\gamma_{\mathrm{s}}$ replaced by $\sigma^2$], so the asymptotic floor of any parallel protocol lies between $\sigma^2$ and the GHZ value $e\sigma^2$. Jump components are subquadratic in the order-$N$ GHZ decoherence rate and hence vanish after division by $N^2$. The convergence rate depends on $\nu$. The jump contribution to $\delta\omega^2T$ is $O(N^{-2})$ for finite activity, while for an $\alpha$-stable phase process ($\Gamma_N\propto N^\alpha$, $\alpha<2$, infinite activity) one finds $\delta\omega^2T\propto N^{\alpha-2}$ and $\delta\omega\propto N^{\alpha/2-1}$. This partial restoration lies between the standard-deviation floor and the Heisenberg limit. Equation~\eqref{eq:Slevyfloor} therefore quantifies the Brownian floor of this GHZ protocol. \emph{Numerical checks:} the pointwise inequality was verified on a dense grid ($q\le999$); the expectation form on $2\times10^4$ random discrete laws with no violation; and the $\alpha$-stable limit $\Gamma_q\propto q^\alpha$, $\Gamma_q/q^2\to0$, for $\alpha\in\{0.5,1,1.5,1.9\}$ by direct quadrature. Figure~\ref{fig:Sbenchmark} shows the finite-time benchmark of Eq.~\eqref{eq:SfiniteT}, the normalized order-$N$ rates, and the resulting local scaling exponents for these noise classes.

\section{A converse bound for parallel strategies: scaling optimality of GHZ}
\label{sec:Sconverse}
\emph{Reduction lemma.} Let the parameter-independent input be an arbitrary pure state $|\Psi\rangle\in(\mathbb C^2)^{\otimes N}\otimes\mathcal H_A$ with a noiseless ancilla, and decompose it as $|\Psi\rangle=\sum_m\sqrt{p_m}|v_m\rangle$, where $|v_m\rangle\in\mathcal H_m\otimes\mathcal H_A$ is normalized and $\mathcal H_m$ is the register's $J_z$ eigenspace. The signal and every kick act as scalars on each such eigenspace. The output is therefore supported on the orthonormal set $\{|v_m\rangle\}$ and has matrix elements
\begin{equation}
 \langle v_m|\rho_\omega|v_{m'}\rangle
 =\sqrt{p_mp_{m'}}\exp\!\left[-i\omega t(m-m')
       +\Gamma t\{\varphi(m-m')-1\}\right].
 \label{eq:Sreduction}
\end{equation}
An $\omega$-independent isometry maps this family to the register-only Dicke-basis input $|\psi_{\bm p}\rangle=\sum_m\sqrt{p_m}|J,m\rangle$. The output QFI depends only on the $N+1$ probabilities $p_m$, not on the ancilla or the degeneracy within an eigenspace. Mixed inputs are covered by purification. Thus ancillas do not increase the optimum, and the input search reduces to the $N+1$-outcome probability simplex. A full-space check for $N=3$ with a four-dimensional ancilla agrees with the reduced QFI to numerical precision.

\emph{Theorem and regularity assumptions.} Let the kick density have finite location Fisher information in the weak-derivative sense:
\begin{equation}
 \sqrt p\in H^1(\mathbb R),\qquad
 J[p]=4\int_{\mathbb R}\!|\partial_a\sqrt{p(a)}|^2\,\dd a
     =\int_{\{p>0\}}\!\frac{[p'(a)]^2}{p(a)}\,\dd a<\infty .
 \label{eq:SregularFI}
\end{equation}
Here $H^1$ requires both the function and its weak derivative to be square integrable. This excludes a density that is discontinuous at a support boundary even if its ordinary derivative vanishes almost everywhere. The Laplace cusp is allowed and gives $J[p]=1/\mu^2$.

For $\Gamma>0$, every protocol comprising an arbitrary ancilla-assisted input, one uninterrupted parallel interrogation of all $N$ probes, an arbitrary final POVM, and independent repetitions obeys
\begin{equation}
 \delta\omega^2T\ge
 \left[\frac{N^2}{e\Gamma}+\frac{cJ[p]}{\Gamma}\right]^{-1},
 \qquad
 c=\sup_{x>0}x\sum_{k\ge1}e^{-x}\frac{x^k}{k!\,k}
   =1.295248\ldots .
 \label{eq:Sconversebound}
\end{equation}
The protocol contains no intermediate control, feedback, or bath-record readout.

\emph{Proof.} Condition mathematically on the Poisson count $K$ in a cycle. Its weights $w_k=e^{-\Gamma t}(\Gamma t)^k/k!$ are independent of $\omega$, so convexity gives $F(t)\le\sum_k w_kF_k(t)$. On the zero-kick branch, the generator range of $tJ_z$ gives $F_0\le t^2N^2$. For $k\ge1$, commutativity of the kicks and signal gives
\begin{equation}
 \rho_k(\omega)=\int\dd s\,p^{*k}(s-\omega t)\,
 U(s)\rho_{\rm in}U^\dagger(s),
 \qquad U(s)=e^{-isJ_z}\otimes\mathbb{I}_A .
 \label{eq:Skicklocation}
\end{equation}
This is a classical location family followed by a parameter-independent preparation map. Data processing and Stam's convolution inequality~\cite{S_Stam} give $F_k\le t^2J[p^{*k}]\le t^2J[p]/k$. Wrapping the phase modulo $2\pi$ is another parameter-independent map and cannot increase this information. Hence
\begin{equation}
 \frac{F(t)}t
 \le t e^{-\Gamma t}N^2+tJ[p]\sum_{k\ge1}\frac{w_k}{k}
 \le\frac{N^2}{e\Gamma}+\frac{cJ[p]}{\Gamma}.
 \label{eq:Sconverseproof}
\end{equation}
Taking the reciprocal of the optimized information rate proves Eq.~\eqref{eq:Sconversebound}. Restricting the optimization to $t\le T$ can only strengthen the bound. The conditioning is an upper-bound construction; it does not presume that the experimentalist can identify and retain zero-kick runs. $\square$

For reproducibility, the scalar function defining $c$ can be evaluated as
\begin{equation}
 h(x)=xe^{-x}\left[\operatorname{Ei}(x)-\gamma_{\rm E}-\log x\right],
 \qquad x_{\rm max}=4.168485\ldots ,
 \label{eq:Scconstant}
\end{equation}
where $\gamma_{\rm E}$ is Euler's constant. If $g(x)=e^xh'(x)$, then $g''(x)=(1+x-e^x)/x^2<0$; $g$ is initially positive and eventually negative, giving a unique positive stationary point of $h$. Direct maximization yields the stated $c$.

For Laplace kicks the bound becomes
$\delta\omega\sqrt T\ge(\sqrt{e\Gamma}/N)
[1+ce/(\mu N)^2]^{-1/2}$.
The GHZ value approaches $\sqrt{2e\Gamma}/N$, establishing its optimal $1/N$ exponent within this protocol class while leaving a factor $\sqrt2$ in the leading standard-deviation constant. In the fixed-$D_2$ diffusion limit at fixed $N$, $\Gamma\to\infty$ and $\mu^2=D_2/(2\Gamma)$, the bound tends to $\delta\omega^2T\ge D_2/(2c)$, with Gaussian $N^0$ scaling.

\emph{Bath records and the factor $\sqrt2$.} The proof conditions on the count $K$, so Eq.~\eqref{eq:Sconversebound} also holds for protocols that read a classical record of the event times after a fixed interrogation: a known label leaves the bound $F\le\sum_kw_kF_k$ unchanged. A GHZ protocol that discards kicked cycles retains the information $N^2t^2$ with probability $e^{-\Gamma t}$, giving the rate $N^2te^{-\Gamma t}$, which is maximal at $t=1/\Gamma$ and equals the $N^2$ term $N^2/(e\Gamma)$ of Eq.~\eqref{eq:Sconverseproof}. Without the record, the unconditional GHZ rate is $N^2/(2e\Gamma_N)\to N^2/(2e\Gamma)$; the factor $2$ in information, $\sqrt2$ in standard deviation, is the information lost when the bath is not monitored. A record available in real time allows a cycle to end at the first kick. With the cycle duration $\min(\tau_1,t)$, where $\tau_1$ is the time of the first kick, the information rate becomes $N^2t^2e^{-\Gamma t}\Gamma/(1-e^{-\Gamma t})$, which is maximal near $\Gamma t\simeq1.6$ with the value $0.648\,N^2/\Gamma$. This exceeds the $N^2$ term of Eq.~\eqref{eq:Sconverseproof} and is consistent with the theorem, whose protocol class requires one uninterrupted interrogation of fixed duration.

\emph{Symmetric weights in the numerical search.} Write $\mathcal F_t(\bm p)$ for the QFI of Eq.~\eqref{eq:Sreduction} at fixed $t$. Prepare a coherent superposition of $|\psi_{\bm p}\rangle|0\rangle_A$ and $|\psi_{\bm r}\rangle|1\rangle_A$ with weights $\lambda$ and $1-\lambda$. The reduction lemma gives the QFI $\mathcal F_t(\lambda\bm p+(1-\lambda)\bm r)$. Measuring the ancilla label produces the two corresponding output families with parameter-independent probabilities. QFI monotonicity therefore implies
\begin{equation}
 \mathcal F_t(\lambda\bm p+(1-\lambda)\bm r)
 \ge\lambda\mathcal F_t(\bm p)+(1-\lambda)\mathcal F_t(\bm r).
 \label{eq:Sweightconcavity}
\end{equation}
For the symmetric Laplace channel, reflecting $m\to-m$ leaves the QFI unchanged. Averaging a weight vector with its reflection consequently does not lower the QFI. This justifies restricting the fixed-$t$ optimum to symmetric weights; it does not by itself certify a numerical optimization over $t$.

\emph{Finite-$N$ search and probe-family comparison.} The MATLAB search uses $N\in\{2,4,6,8,10\}$, $\mu\in\{0.2,0.5\}$, a finite time grid, and multistart Nelder--Mead optimization. Below the crossover it finds interior-peaked weight profiles with an information-rate advantage over GHZ, up to approximately $36\%$ at $N=2$, $\mu=0.2$. This is consistent with constant-factor improvements over GHZ under Gaussian collective dephasing~\cite{Dorner2012}. At $\mu N=2$ the gain is approximately $0.2\%$ for $N=10$, $\mu=0.2$, and no improvement is resolved at the tested points with $\mu N\ge3$.

The circles in main-text Fig.~2(a) use a separate Python search at $\mu=0.1$, $\gamma_{\mathrm{s}}=1$, for
$N=2,\ldots,8$, $10$, $12$, $14$, $16$, $20$, $24$, and $30$.
Symmetric weights and $\log t$ are optimized with multistart L-BFGS-B, a logarithmic time scan for initialization, and bounded time refinement. The information-rate gains are $48.16\%$ at $N=2$, $76.93\%$ at $N=4$, $4.15\%$ at $N=12$, $1.85\%$ at $N=14$, and $0.32\%$ at $N=20$; none is resolved at $N=30$. The complete values accompany the figure data. All fourteen points satisfy Eq.~\eqref{eq:Sconversebound}; the information upper bound divided by the achieved information rate ranges from $2.50$ to $4.62$.

\begin{figure}[ht]
\centering
\subfloat[]{\includegraphics{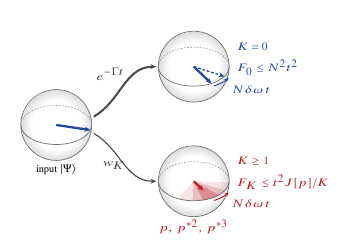}}\hspace{3pt}%
\subfloat[]{\includegraphics{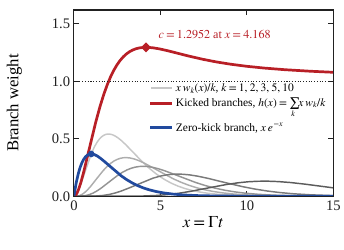}}\hspace{3pt}%
\subfloat[]{\includegraphics{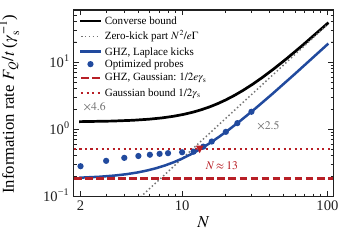}}
\caption{\label{fig:Sconverse} Converse bound for parallel strategies ($\mu=0.1$, $\gamma_{\mathrm{s}}=1$). (a)~Geometric scheme of Eqs.~\eqref{eq:Sconversebound}--\eqref{eq:Sconverseproof} on the effective Bloch sphere. Without kicks ($K=0$, weight $e^{-\Gamma t}$) the phase remains well defined and $\delta\omega$ rotates it by $N\delta\omega\,t$, so the information is bounded by $N^2t^2$. After $K\ge1$ kicks (weights $w_K$) the phase is spread over the convolved kick law $p^{*K}$ (wedges for $K=1,2,3$) and $\delta\omega$ only shifts it, so the information is at most $t^2J[p]/K$. (b)~Zero-kick weight $xe^{-x}$, kicked-branch terms $x\,w_k(x)/k$ for $k=1,2,3,5,10$, and their sum $h(x)$ of Eq.~\eqref{eq:Scconstant}, maximal at $x=4.168$ with $h=c=1.2952$. (c)~Information rates versus $N$: the converse bound and its zero-kick part, the GHZ rate, the optimized probes of main-text Fig.~2(a) (circles), and the fixed-$T_2$ Gaussian GHZ rate and input-independent bound of Eq.~\eqref{eq:Sgaussianparallel}, which the GHZ rate crosses at $N\approx13$.}
\end{figure}

Neither search certifies a global optimum; the two agree around the crossover variable $\mu N$. Figure~\ref{fig:Sconverse} sketches the conditioning argument, shows the two branch functions of the proof, and compares the bound with the achieved information rates. The exact ordering within the two-branch Dicke-cat family is derived in Sec.~S7 and complements these multi-level searches.

\section{Independent dephasing, amplitude damping, and a Brownian share}
\label{sec:Srobustness}
Let $\Gamma_N$ denote the common order-$N$ dephasing rate, including any Brownian contribution when present. Adding per-atom dephasing $\frac12\gamma_{\mathrm{i}}\sum_i\mathcal D[\sigma_z^{(i)}]$ multiplies the GHZ coherence by $e^{-N\gamma_{\mathrm{i}}t}$. For amplitude damping $\gamma_\downarrow\sum_i\mathcal D[\sigma_-^{(i)}]$, with $\sigma_-|e\rangle=|g\rangle$, the additional factor is $e^{-N\gamma_\downarrow t/2}$. In the notation of main-text Eq.~(9), the coherence rate is
\begin{equation}
 \Gamma_{\mathrm{coh}}=\Gamma_N+N\left(\gamma_{\mathrm{i}}+\frac{\gamma_\downarrow}{2}\right).
 \label{eq:Srobust}
\end{equation}
Pure dephasing preserves the GHZ endpoint populations, whereas amplitude damping transfers population into intermediate excitation sectors. The latter requires a block-resolved QFI.

\emph{Amplitude-damping QFI.} Write $\eta=e^{-\gamma_\downarrow t}$. In the basis $(|e^N\rangle,|g^N\rangle)$, the only $\omega$-dependent block is
\begin{equation}
 \rho_{\rm end}(t)=
 \begin{pmatrix}a&c_\omega\\c_\omega^*&b\end{pmatrix},
 \quad a=\frac{\eta^N}{2},\quad
 b=\frac{1+(1-\eta)^N}{2},\quad
 |c_\omega|=\frac{\eta^{N/2}}2e^{-(\Gamma_N+N\gamma_{\mathrm{i}})t}.
 \label{eq:Sendblock}
\end{equation}
The coherence carries $e^{-iN\omega t}$ and any calibrated asymmetric-kick phase. All other blocks are diagonal and independent of $\omega$. Their weights, including $s=a+b$, are also independent of $\omega$. Consequently the full QFI is $s$ times the QFI of the normalized endpoint block:
\begin{equation}
 F_Q^{(T_1)}(t)=\frac{4N^2t^2|c_\omega|^2}{s}
 =\frac{N^2t^2\eta^Ne^{-2(\Gamma_N+N\gamma_{\mathrm{i}})t}}
 {[1+\eta^N+(1-\eta)^N]/2}.
 \label{eq:ST1QFI}
\end{equation}
At $\gamma_\downarrow=0$, this reduces to Eq.~\eqref{eq:SFQ} with $\Gamma_N\mapsto\Gamma_N+N\gamma_{\mathrm{i}}$. Independent dephasing alone therefore gives $\delta\omega^2T=2e(\Gamma_N+N\gamma_{\mathrm{i}})/N^2$ when its interior time optimum is allowed. With amplitude damping, the quantity to maximize is Eq.~\eqref{eq:ST1QFI}/$t$.

For pure $T_1$ noise at large $N$, put $x=N\gamma_\downarrow t$. Then
$F_Q^{(T_1)}/t\to(2N/\gamma_\downarrow)x/(e^x+1)$, which gives
\begin{equation}
 \max_t\frac{F_Q^{(T_1)}}t
 =0.556929\ldots\frac{N}{\gamma_\downarrow},\qquad
 x^*=1+W(e^{-1})=1.278465\ldots .
 \label{eq:ST1opt}
\end{equation}
Thus $\delta\omega^2T=1.795561\ldots\,\gamma_\downarrow/N$, with a different prefactor from a dephasing-only substitution into Eq.~\eqref{eq:Sdw}. The no-amplitude-damping limit and the endpoint-block formula are checked independently by full Hilbert-space Kraus propagation in Sec.~S13.

With pure finite-activity common noise, the $1/N$ standard-deviation regime requires
$N(\gamma_{\mathrm{i}}+\gamma_\downarrow/2)\ll\Gamma_N\simeq\Gamma$.
Independent noise eventually produces SQL scaling. The scale
$N\sim\Gamma/(\gamma_{\mathrm{i}}+\gamma_\downarrow/2)$ is an order-of-magnitude crossover; with $\gamma_\downarrow>0$ the QFI of Eq.~\eqref{eq:ST1QFI} must be optimized.

\emph{Mixed common noise at fixed $T_2$.} For finite-activity jumps plus Brownian phase diffusion, retain $\gamma_{\mathrm{s}}$ as the total common order-one rate:
\begin{equation}
 \Gamma_q^{\rm common}=\frac{\sigma^2q^2}{2}
      +\Gamma[1-\operatorname{Re}\varphi(q)],
 \qquad
 \gamma_{\mathrm{s}}=\frac{\sigma^2}{2}
      +\Gamma[1-\operatorname{Re}\varphi(1)] .
 \label{eq:Smixedcalibration}
\end{equation}
If $f=\sigma^2/(2\gamma_{\mathrm{s}})$ is the Brownian share, the Laplace jump rate is
$\Gamma=(1-f)\gamma_{\mathrm{s}}(1+\mu^2)/\mu^2$.
The $1\%$ curve in main-text Fig.~2(b) has $f=0.01$, so its jump contribution at order one is $0.99\gamma_{\mathrm{s}}$; the Brownian share is part of the fixed $\gamma_{\mathrm{s}}$, not an addition to it.

For independent pure dephasing and an allowed interior optimum, writing
$\Gamma_N^{\rm jump}=\Gamma[1-\operatorname{Re}\varphi(N)]$ yields
\begin{equation}
 \delta\omega^2T=e\sigma^2+
       \frac{2e\Gamma_N^{\rm jump}}{N^2}
       +\frac{2e\gamma_{\mathrm{i}}}{N}.
 \label{eq:Scombinedprecision}
\end{equation}
In the Laplace saturation regime, a well-separated Heisenberg window therefore requires
\begin{equation}
 \frac1\mu\ll N\ll
 \min\!\left\{\frac{\Gamma}{\gamma_{\mathrm{i}}},
              \sqrt{\frac{2\Gamma}{\sigma^2}}\right\}.
 \label{eq:Scombinedwindow}
\end{equation}
A vanishing noise share removes its corresponding upper cutoff. Brownian diffusion restores a floor; independent dephasing gives an SQL tail. Equation~\eqref{eq:Scombinedprecision} is for pure dephasing; with amplitude damping, Eq.~\eqref{eq:ST1QFI} applies.

\section{Dicke-cat family and the scope of probe optimality}
\label{sec:Sprobe}
The converse in Sec.~S5 proves the GHZ scaling exponent within the specified parallel protocol class. A separate, exact ordering is available within the Dicke-cat family. Let
$|\psi_q\rangle=(|J,q/2\rangle+|J,-q/2\rangle)/\sqrt2$, where $0<q\le N$ has the same parity as $N$ so that both states lie on the Dicke ladder. The signal phase is $q\omega t$ and the decay rate is $\Gamma_q$, hence $F_Q=q^2t^2e^{-2\Gamma_qt}$. For symmetric Laplace kicks and an interior optimum,
\begin{equation}
 \delta\omega^2T=\frac{2e\Gamma_q}{q^2}
 =\frac{2e\Gamma\mu^2}{1+\mu^2q^2},
 \label{eq:Scatprecision}
\end{equation}
which strictly decreases with the allowed $q$. GHZ, $q=N$, is thus exactly optimal within this family.

The ordering also survives the finite-time constraint. On the branch with $T\le(2\Gamma_q)^{-1}$, use Eq.~\eqref{eq:SfiniteT} with $N$ replaced by $q$. Treating $q$ as a continuous interpolation for comparison,
\begin{equation}
 \frac{\dd\log\mathcal V_q(T)}{\dd q}
 =\frac{4\Gamma\mu^2qT}{(1+\mu^2q^2)^2}-\frac2q
 \le-\frac{2\mu^2q}{1+\mu^2q^2}<0 .
 \label{eq:ScatfiniteT}
\end{equation}
The constrained and interior branches join continuously. The maximum allowed order minimizes the benchmark for any $T>0$ under this Laplace model.

\begin{figure}[ht]
\centering
\subfloat[]{\includegraphics{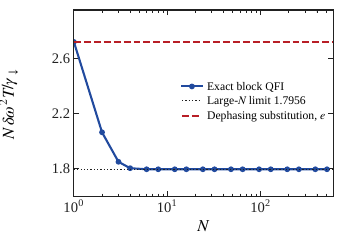}}\hspace{3pt}%
\subfloat[]{\includegraphics{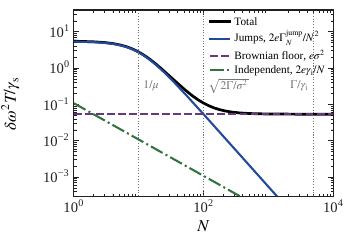}}\hspace{3pt}%
\subfloat[]{\includegraphics{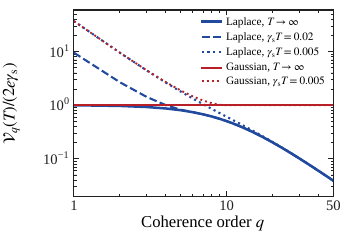}}
\caption{\label{fig:Srobust} Residual noise and the Dicke-cat family. (a)~Amplitude damping: optimized $N\,\delta\omega^2T/\gamma_\downarrow$ from the exact block QFI, Eq.~\eqref{eq:ST1QFI}, approaching $1.7956$ (dotted); the dephasing substitution $N\gamma_\downarrow/2$ in Eq.~\eqref{eq:Sdw} gives $e$ (dashed). (b)~Terms of Eq.~\eqref{eq:Scombinedprecision} for the parameters of main-text Fig.~2(b) ($\mu=0.1$, $\gamma_{\mathrm{i}}=0.02\gamma_{\mathrm{s}}$, $f=0.01$) and their sum; dotted verticals mark the window limits $1/\mu$, $\sqrt{2\Gamma/\sigma^2}$, and $\Gamma/\gamma_{\mathrm{i}}$. (c)~Dicke-cat benchmark $\mathcal V_q(T)/(2e\gamma_{\mathrm{s}})$, Eqs.~\eqref{eq:Scatprecision}--\eqref{eq:ScatfiniteT}, for Laplace kicks ($\mu=0.1$) at $T\to\infty$ and two finite budgets, with the Gaussian values (red); it decreases with $q$ on both branches.}
\end{figure}

For Gaussian common dephasing, the interior benchmark is $2e\gamma_{\mathrm{s}}$ for every $q$. General multi-level probes can improve the GHZ prefactor while retaining $N^0$ scaling~\cite{Dorner2012}; the finite-$N$ searches in Sec.~S5 are distinct from the exact two-branch calculation. Neither Eq.~\eqref{eq:Scatprecision} nor the converse proves global finite-$N$ optimality of GHZ or equality of its asymptotic prefactor with the converse. Under independent dephasing or amplitude damping, a general Dicke cat is not characterized by $\Gamma_q$ alone; the family calculation here is for pure common dephasing. Figure~\ref{fig:Srobust} illustrates the exact amplitude-damping prefactor, the decomposition of Eq.~\eqref{eq:Scombinedprecision}, and the monotone Dicke-cat ordering.

\FloatBarrier
\section{Prospective atomic-clock parameter estimates}
\label{sec:Sclock}
We consider local estimation of a constant detuning $\omega=\omega_{\rm atom}-\omega_{\rm LO}$, with LO fluctuations entering as common phase noise through $J_z$~\cite{S_Andre2004,S_Kessler2014,S_Schulte2020,S_Ludlow2015}. Ideal instantaneous LO phase jumps $\beta$ at rate $\Gamma$ realize $e^{-i\beta J_z}$, with $\beta$ distributed according to the Laplace law and $\mu$ its phase scale. The comparison is between local GHZ Ramsey precisions with separately optimized interrogation times and does not model a feedback-stabilized clock.

The illustrative values $\Gamma=1~\mathrm{s}^{-1}$, $\mu=0.2$, and $\gamma_{\mathrm{i}}=10^{-3}~\mathrm{s}^{-1}$ give
$\gamma_{\mathrm{s}}=\Gamma\mu^2/(1+\mu^2)=\Gamma/26=0.0385~\mathrm{s}^{-1}$.
The common-noise coherence time is $T_2=26$ s; with independent dephasing, the observable single-atom total coherence time is
$T_{2,\rm tot}=(\gamma_{\mathrm{s}}+\gamma_{\mathrm{i}})^{-1}=25.3$ s in both compared models. With no Brownian share, the crossover interval is approximately $5\lesssim N\lesssim10^3$.

\begin{figure}[ht]
\centering
\subfloat[]{\includegraphics{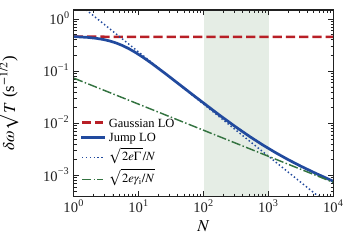}}\hspace{3pt}%
\subfloat[]{\includegraphics{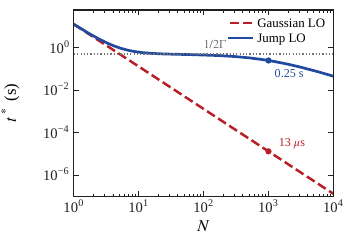}}\hspace{3pt}%
\subfloat[]{\includegraphics{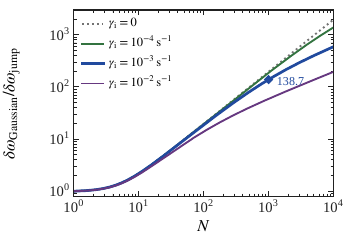}}
\caption{\label{fig:Sclock} Idealized GHZ Ramsey comparison at fixed common-noise $T_2$ ($\Gamma=1~\mathrm{s}^{-1}$, $\mu=0.2$, $\gamma_{\mathrm{i}}=10^{-3}~\mathrm{s}^{-1}$; logarithmic axes). (a)~Optimized precision for the jump and Gaussian local oscillators (LO) with the same independent dephasing; thin lines: the asymptotes $\sqrt{2e\Gamma}/N$ and $\sqrt{2e\gamma_{\mathrm{i}}/N}$; shading: the $N=10^2$--$10^3$ crossover interval. (b)~Optimal interrogation times $t^*=1/(2\Gamma_{\mathrm{coh}})$; markers: the $N=10^3$ values of Eq.~\eqref{eq:Sclocktimes}. (c)~Precision ratio for $\gamma_{\mathrm{i}}=10^{-4}$, $10^{-3}$, $10^{-2}~\mathrm{s}^{-1}$ and $\gamma_{\mathrm{i}}=0$ (dotted); marker: the value $138.7$ of Eq.~\eqref{eq:Sclockratio}.}
\end{figure}

Both curves use GHZ probes and optimize their interrogation times separately, retaining the same independent rate; Fig.~\ref{fig:Sclock} shows the two precision curves, their optimal interrogation times, and the precision ratio for several independent rates. Their local precision ratio is
\begin{equation}
 \frac{\delta\omega_{\rm Gaussian}}{\delta\omega_{\rm jump}}
 =\sqrt{\frac{\gamma_{\mathrm{s}}N^2+N\gamma_{\mathrm{i}}}
              {\Gamma_N+N\gamma_{\mathrm{i}}}}
 =138.7\quad(N=10^3).
 \label{eq:Sclockratio}
\end{equation}
At this atom number, the corresponding times are
\begin{equation}
 t^*_{\rm jump}=\frac1{2(\Gamma_N+N\gamma_{\mathrm{i}})}
 =0.250~\mathrm{s},\qquad
 t^*_{\rm Gaussian}=\frac1{2(\gamma_{\mathrm{s}}N^2+N\gamma_{\mathrm{i}})}
 =13.0~\mu\mathrm{s}.
 \label{eq:Sclocktimes}
\end{equation}
The estimate $1/(2\Gamma)=0.5$ s applies to saturated common kicks alone and does not include the independent-noise contribution at $N=10^3$.

In the saturation regime $\Gamma_N\simeq\Gamma$ and for $\mu N\gg1$, Eq.~\eqref{eq:Sclockratio} reduces to
\begin{equation}
 \frac{\delta\omega_{\rm Gaussian}}{\delta\omega_{\rm jump}}
 \simeq\mu N\left(1+\frac{N\gamma_{\mathrm{i}}}{\Gamma}\right)^{-1/2},
 \label{eq:Sclockclosed}
\end{equation}
so the gain over a Gaussian LO with the same $T_2$ is $\mu N$ inside the window and is reduced by $\sqrt2$ at the crossover $N=\Gamma/\gamma_{\mathrm{i}}$; for the values above, Eq.~\eqref{eq:Sclockclosed} gives $141$ at $N=10^3$, within 2\% of Eq.~\eqref{eq:Sclockratio}. With a Brownian share $f$ of the common single-atom rate [Eq.~\eqref{eq:Smixedcalibration}], the ratio saturates at $1/\sqrt f$ for $N\gg\sqrt{2\Gamma/\sigma^2}$, which is $10$ for $f=0.01$.

Equation~\eqref{eq:Sclockratio} assumes the calibrated noise and local-fringe conditions of Sec.~S3, independent increments, negligible cycle overhead, and no Brownian component. Real LO spectra may include colored thermal or flicker-frequency noise~\cite{S_Schulte2020}, which is not automatically the Wiener phase process in Sec.~S4. A justified Brownian share imposes the cutoff in Eq.~\eqref{eq:Scombinedwindow}.

\FloatBarrier

\section{Relation to metrological bounds}
\label{sec:Sbounds}
Independent-noise precision bounds apply to the residual per-particle channels in Sec.~S6~\cite{S_Escher2011,S_Demkowicz2012,S_Kolodynski2013}. The common event process acts on the register as a whole and is not a tensor product of single-atom channels; its no-kick probability $e^{-\Gamma t}$ does not decay with the number of atoms. Thus the usual independent-channel assumptions cannot be imposed by substituting the same measured single-atom $T_2$. The Hamiltonian-not-in-Lindblad-span criterion addresses a complementary setting with fast control and noiseless ancillas~\cite{S_Zhou2018}. The gain established here is in atom-number scaling; the repeated Ramsey protocol accumulates information linearly in $T$.

The Gaussian floor quoted in Sec.~S3 is the GHZ benchmark. An input-independent comparison follows by representing Gaussian common dephasing as a random phase with variance $2\gamma_{\mathrm{s}}t$ and mean $\omega t$. Its classical location Fisher information bounds the QFI of every output obtained from a parameter-independent preparation map:
\begin{equation}
 F_Q(t)\le\frac{t}{2\gamma_{\mathrm{s}}},\qquad
 \delta\omega^2T\ge2\gamma_{\mathrm{s}} .
 \label{eq:Sgaussianparallel}
\end{equation}
GHZ attains $2e\gamma_{\mathrm{s}}$ when its interior time optimum is allowed, so arbitrary-probe improvements can change the constant while leaving the $N^0$ scaling [Fig.~\ref{fig:Sconverse}(c)]. This is consistent with the collective-dephasing analysis of Ref.~\cite{Dorner2012}. In the finite-activity case, Sec.~S3 gives an achievable $N^2$ information rate, and Sec.~S5 supplies the matching exponent bound.

General correlated-noise bounds also cover adaptive channel-use models~\cite{S_Kurdzialek2025}. Their applicability and tightness depend on the actual correlation model and access to controls; the converse here applies to the uninterrupted parallel class defined in Sec.~S5. Noise not parallel to the signal changes the generator geometry and permits superclassical scaling even with Markovian Gaussian statistics~\cite{S_Chaves2013,S_Brask2015}. Here the coupling stays parallel; what changes is the high-order phase statistics.

\emph{Echo sequences.} Because the kick process is white, sequences of collective $\pi$ pulses do not change its effect on a coherence. In the toggling frame of $\pi$ pulses about an equatorial axis, a kick $e^{-iaJ_z}$ applied after an odd number of pulses enters as $e^{+iaJ_z}$, so the accumulated phase is $\Theta=\sum_is_ia_i$ with signs $s_i=\pm1$ fixed by the sequence. For a symmetric kick law, $s_ia_i$ has the law of $a_i$, and $\Theta$ has the same distribution as without pulses; the coherence factor $e^{-\Gamma_qt}$ is unchanged. Within the protocols considered here, the event record of Sec.~S5 is therefore the only additional resource against white common kicks.

\section{A uniform finite-time bound for finite-activity noise}
\label{sec:Sfinact}
The compound-Poisson bath of the main text has independent, exponentially distributed waiting times. A weaker statement remains true without either assumption. Let the accumulated common phase $\Theta(t)$ change only at the epochs of a stationary marked point process of mean rate $R$, with arbitrary arrival correlations and arbitrary (possibly correlated) marks, and let $\Theta$ be constant between events with $\Theta(0)=0$. Conditioning only on whether the event count $K(t)$ vanishes gives
\begin{equation}
 C_q(t)\equiv\mathbb{E}\big[e^{-iq\Theta(t)}\big]=p_0(t)+\big[1-p_0(t)\big]\chi_q(t),\qquad |\chi_q(t)|\le1,\qquad p_0(t)=\mathbb{P}[K(t)=0]\ge1-Rt,
\end{equation}
where the last step is Markov's inequality, $\mathbb{P}[K\ge1]\le\mathbb{E}[K]=Rt$. Hence, for every coherence order $q$ and every interval,
\begin{equation}
1-\big|C_q(t)\big|\le2\big[1-p_0(t)\big]\le2Rt:
\label{eq:Sfinact}
\end{equation}
the finite-time coherence loss is bounded uniformly in $q$ by twice the expected number of events. Here $C_q$ is the complex coherence factor and the Ramsey contrast is $|C_q|$. Equation~\eqref{eq:Sfinact} is a finite-time inequality; exponential decay, a time-independent rate, and a large-$q$ limit require the conditions given below.

A sufficient condition gives a sharper finite-window statement. If the marks are iid, independent of the count, and have characteristic function $\varphi(q)$, then
\begin{equation}
C_q(t)=\mathbb{E}\!\left[\varphi(q)^{K(t)}\right]\equiv G_t\!\left(\varphi(q)\right),
\label{eq:Spgf}
\end{equation}
where $G_t(z)=\mathbb{E}[z^{K(t)}]$ is the probability-generating function of the finite-window count. If the mark law is Rajchman---in particular, if it has an absolutely continuous density, so that $\varphi(q)\to0$ by the Riemann--Lebesgue lemma---dominated convergence gives
\begin{equation}
C_q(t)\xrightarrow[|q|\to\infty]{}G_t(0)=\mathbb{P}[K(t)=0]\equiv p_0(t).
\label{eq:Svoidlimit}
\end{equation}
Thus, when $p_0(t)>0$, the finite-window exponent $\Lambda_q(t)\equiv-t^{-1}\log|C_q(t)|$ approaches $-t^{-1}\log p_0(t)$. It equals the constant $R$ for the void law $p_0(t)=e^{-Rt}$; the void law alone does not establish all the independence properties of a Poisson process. More generally, Eq.~\eqref{eq:Svoidlimit} remains true if the conditional characteristic function of the total phase given $K(t)=k$ tends to zero for every $k\ge1$; otherwise only the uniform bound~\eqref{eq:Sfinact} is established. For the compound-Poisson model, independent increments give the semigroup and the constant rate. For a general correlated point process, applying the $T/t$ information accumulation of Sec.~S3 would additionally require independent cycles, a reset, or an explicit analysis of inter-cycle correlations.

\FloatBarrier

\section{Dissipative cascade and the superradiance cap}
\label{sec:Scascade}
This section develops a dissipative analogue of the finite-event-rate mechanism, separate from the random-unitary dephasing channel used for frequency estimation. We extend the emission sector of Ref.~\cite{Funo2024} to the full Dicke ladder and compare it with conventional collective emission~\cite{S_Dicke1954,S_GrossHaroche1982,S_Bohnet2012,S_Norcia2016}.

\emph{Single-event channel.} A zero-temperature collision exchanges at most one excitation with the register. For $L=J^-$ and $Q=J^+J^-$, its two Kraus operators can be written
\begin{equation}
 L_a=-iJ^-\frac{\sin(a\sqrt Q)}{\sqrt Q},
 \qquad C_a=\cos(a\sqrt Q),\qquad
 \mathcal E_a(\rho)=L_a\rho L_a^\dagger+C_a\rho C_a .
 \label{eq:Scollisionmap}
\end{equation}
The ratio $\sin(a\sqrt x)/\sqrt x$ is defined by its limit $a$ at $x=0$.
Since $L_a^\dagger L_a=\sin^2(a\sqrt Q)$, one has
$L_a^\dagger L_a+C_a^\dagger C_a=\mathbb{I}$.
The map is therefore completely positive and trace preserving. With $M_a=C_a-\mathbb{I}$,
$\mathcal E_a-\mathbb{I}=\mathcal D[L_a]+\mathcal D[M_a]$ as superoperators.

We average over the one-sided pulse-area density
\begin{equation}
 p_+(a)=\mu^{-1}e^{-a/\mu}\Theta(a),\qquad
 \langle a^2\rangle_{p_+}=2\mu^2 ,
 \label{eq:Semissiondensity}
\end{equation}
which is distinct from the symmetric Laplace law used for dephasing. Poisson collisions give the dissipative generator
\begin{equation}
 \dot\rho=\Gamma\int_0^\infty\dd a\,p_+(a)
       [\mathcal E_a(\rho)-\rho]
 =\Gamma\int_0^\infty\dd a\,p_+(a)
       \{\mathcal D[L_a]+\mathcal D[M_a]\}\rho .
 \label{eq:Semissiongenerator}
\end{equation}
A Hamiltonian diagonal in $J_z$ does not change the populations considered below. For weak pulses, $L_a=-iaJ^-+O(a^3)$ and $\mathcal D[M_a]=O(a^4)$, giving the limiting generator $D_{\rm em}\mathcal D[J^-]$, with $D_{\rm em}=2\Gamma\mu^2$. In this section the comparison is made at fixed weak-pulse emission coefficient $D_{\rm em}$; the dephasing $T_2$ plays no role.

\begin{figure}[ht]
\centering
\subfloat[]{\includegraphics{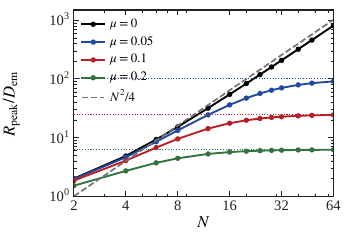}}\hspace{3pt}%
\subfloat[]{\includegraphics{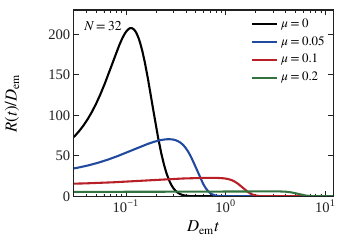}}
\caption{\label{fig:Ssuperradiance} Dissipative analogue: cascade from the fully excited Dicke state. (a)~Peak emission rate versus atom number: Gaussian collective emission grows as $N^2$, finite pulses saturate at $D_{\rm em}/(4\mu^2)$ (dotted). (b)~Emission traces at $N=32$. Rates and times are in units of the weak-pulse coefficient $D_{\rm em}$.}
\end{figure}

\emph{Population dynamics.} A Dicke-state initial condition stays diagonal because $L_a$ lowers one ladder step and $M_a$ is diagonal. Writing $d_m=J(J+1)-m(m-1)$, the downward rate is
\begin{equation}
 W_m=\Gamma\int_0^\infty\dd a\,p_+(a)\sin^2(a\sqrt{d_m})
 =\frac{2\Gamma\mu^2d_m}{1+4\mu^2d_m}
 =\frac{D_{\rm em}d_m}{1+4\mu^2d_m}.
 \label{eq:SWm}
\end{equation}
Here the matrix element of $J^-$ cancels the denominator in $L_a$, and
$\int_0^\infty \mu^{-1}e^{-a/\mu}\sin^2(a\sqrt d)\dd a
 =[1-(1+4\mu^2d)^{-1}]/2$.
The pure-death chain and its emission rate are
\begin{equation}
 \dot P_m=W_{m+1}P_{m+1}-W_mP_m,\qquad
 W_{-J}=0,\qquad
 R(t)=-\frac{\dd\langle J_z\rangle}{\dd t}=\sum_mW_mP_m(t).
 \label{eq:Scascade}
\end{equation}
The absent term above $m=J$ is zero. Figure~\ref{fig:Ssuperradiance} starts from the fully excited state, $P_m(0)=\delta_{m,J}$.

For Gaussian collective emission, $W_m=D_{\rm em}d_m$ and the fully excited cascade has $N^2$ peak scaling. The largest individual rate gives the bound
\begin{equation}
 R_{\rm peak}\le D_{\rm em}\max_m d_m
 =D_{\rm em}\left\lfloor\frac{(N+1)^2}{4}\right\rfloor .
 \label{eq:Sgaussianemissioncap}
\end{equation}
The factor $1/4$ is an upper bound; the population-averaged peak at $N=64$ gives $R_{\rm peak}/(D_{\rm em}N^2)\simeq0.199$. For $\mu>0$, Eq.~\eqref{eq:SWm} implies
\begin{equation}
 0\le W_m<\frac{\Gamma}{2},\qquad
 R(t)\le\frac{\Gamma}{2}=\frac{D_{\rm em}}{4\mu^2},
 \label{eq:Semissioncap}
\end{equation}
with saturation of the active transition rates when $\mu^2d_m\gg1$. This caps the emission independently of atom number. The mid-ladder crossover is of order $N\sim1/\mu$; the peak shape also depends on the initial state and the evolving population distribution.

Projecting the full vectorized generator onto the diagonal subspace reproduces Eq.~\eqref{eq:Scascade}. For $N=20$, $\mu=0.1$, $D_{\rm em}=1$, the maximum difference is $1.2\times10^{-9}$ with 40-point Gauss--Laguerre quadrature.

\FloatBarrier

\section{Coherence-order diagnostics of common-noise records from three platforms}
\label{sec:Sempirical}
The main text identifies the counting statistics of the common noise as the property that bounds the decay of high coherence orders. This section applies the corresponding classical diagnostic to open records from three platforms, an ultrastable laser delivered to a distant optical clock through a fiber network, the magnetic environment of a laboratory, and a superconducting qubit array. For a record $x_j$, $j=1,\ldots,M$, of phase increments over windows of duration $\tau$, the empirical characteristic function $\widehat C(u)=M^{-1}\sum_j e^{-iux_j}$ is the classical counterpart of the coherence factor in Eq.~\eqref{eq:Sreduction}, with the argument $u$ in the role of the coherence order. For increments standardized by their sample standard deviation, the same-variance exponent ratio
\begin{equation}
r(u)=\frac{-\log|\widehat C(u)|}{u^2/2},\qquad u>0,
\label{eq:Sshape}
\end{equation}
equals one at every $u$ for Gaussian increments. A record whose increments contain discrete events has $r(u)<1$ at large $u$, because an event that occurs with bounded probability in a window removes at most a bounded fraction of $|\widehat C(u)|$; this is the classical form of the bound in Eq.~(3) of the main text. These are classical records; the analysis tests the shape of their increment distributions, not the fixed-$T_2$ calibration of Sec.~S4 or the compound-Poisson semigroup itself. Section~S13 describes the data selection, preprocessing, and uncertainty estimation.

\begin{figure}[ht]
\centering
\subfloat[]{\includegraphics{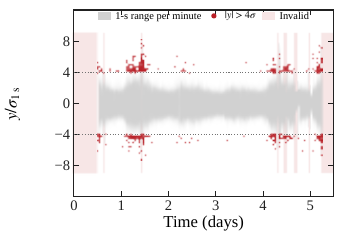}}\hspace{3pt}%
\subfloat[]{\includegraphics{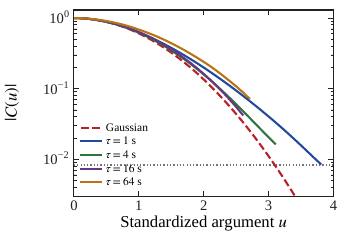}}\hspace{3pt}%
\subfloat[]{\includegraphics{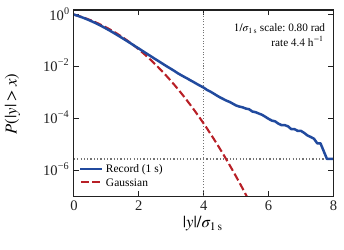}}
\caption{\label{fig:Slaser} One-second record of the transfer beat between the NPL and PTB ultrastable lasers through the 2220-km fiber network~\cite{S_SchioppoData}. (a)~Range of the standardized one-second values in each minute (gray), values beyond $4\sigma$ (red), and invalid stretches (shaded). (b)~Magnitude of the empirical characteristic function of the standardized phase increments over windows of 1, 4, 16, and 64~s against the Gaussian value $e^{-u^2/2}$; each curve ends at its reliability floor $5/\sqrt M$ (dotted line for 1~s). (c)~Survival function of the standardized one-second values against the Gaussian tail; the dotted lines mark the $4\sigma$ threshold and the single-sample level $1/M$.}
\end{figure}

\emph{Ultrastable laser delivered through a fiber network.}
An optical clock interrogates all of its atoms with one laser, so laser phase noise is collective dephasing parallel to the signal. Schioppo \emph{et al.} compared the ultrastable lasers of NPL and PTB through the 2220-km phase-stabilized European fiber network and released the one-second record of the transfer beat between them~\cite{S_Schioppo2022,S_SchioppoData}; the same work evaluates the delivered light as the local oscillator of a distant clock. We analyzed the six-day record of July 2019, 474\,200 consecutive seconds of which 77\% are valid after the authors' flagging of link glitches. At one second the record is dominated by residual noise of the fiber link and not by the two cavities~\cite{S_Schioppo2022}, and its instability of $7.7\times10^{-16}$ corresponds to a phase spread of $0.80$~rad per second at the 194.4-THz carrier. The one-second values have excess kurtosis $1.2$ and a heavy tail [Fig.~\ref{fig:Slaser}(c)]. Values beyond $4\sigma$ occur in 445 clusters, 80\% of which are single seconds, at a mean rate of $4.4$ per hour. Each such value is a phase change of more than $3.2$~rad within one second at the carrier, and the net change over the surrounding seven seconds has median $1.6$~rad. The clusters are not concentrated near flagged gaps, 286 of them lying more than 60~s from any invalid sample, and they carry $3.4\%$ of the variance. Their occurrence is episodic, with hourly counts of Fano factor $9.9$ and no event in $52\%$ of the hours [Fig.~\ref{fig:Slaser}(a)]. The exponent ratio at $u=2$ is $r(2)=0.806$ $[0.784,0.832]$ for 1-s windows and $0.92$, $0.91$, and $0.71$ for 4-, 16-, and 64-s windows, with bootstrap intervals over one-hour blocks [Fig.~\ref{fig:Slaser}(b)]. Removing the largest $1\%$ of the one-second values and restandardizing returns $r(2)=0.91$, so the departure from the Gaussian value is carried by the rare excursions, as for the magnetic records below. For a clock interrogated with this light, an excursion randomizes the phase of every coherence order in the cycle in which it occurs, while the Gaussian part of the record dephases order $q$ at a rate growing as $q^2$. The validity flags of the link constitute a classical event record of the kind considered in Sec.~S5.

\FloatBarrier

\begin{figure}[ht]
\centering
\subfloat[]{\includegraphics{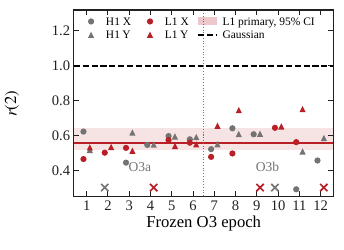}}\hspace{3pt}%
\subfloat[]{\includegraphics{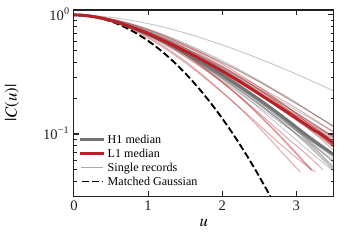}}\hspace{3pt}%
\subfloat[]{\includegraphics{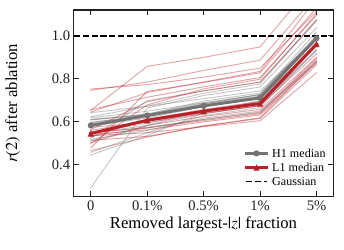}}
\caption{\label{fig:SO3} Preregistered out-of-sample O3 test. (a)~Same-variance exponent ratio $r(2)$ of every usable channel--epoch record (H1 gray, L1 red; X circles, Y triangles); crosses mark unavailable records. Line and band: the L1 primary estimate and its 95\% hierarchical-bootstrap interval. (b)~Empirical characteristic-function magnitudes of all 38 records (thin) and their site medians (bold) against the matched Gaussian. (c)~Tail ablation: $r(2)$ after removing the largest $0.1$--$5\%$ of one-second integrals and restandardizing, for each record (thin) and the site medians (bold).}
\end{figure}

\emph{Magnetic environment of a laboratory.}
Magnetic-field fluctuations shift the Zeeman-sensitive transitions of every atom in a register, so a magnetometer record at a laboratory site becomes a common-phase record once a coupling is chosen. We use the magnetometers of the two LIGO sites, whose records are released through the Gravitational Wave Open Science Center~\cite{S_GWOSC,S_GWOSCO3,S_GWOSCO3Aux}. For an out-of-sample test, twelve disjoint three-hour intervals of the third observing run (O3) and four Hanford/Livingston remote-vault magnetometer axes were frozen before the analysis waveforms were accessed.
After complete-record filtering, each one-second integral series was centered and standardized by its own sample standard deviation. The exponent ratio of Eq.~\eqref{eq:Sshape} was evaluated at $u=2$; a centered Gaussian has $r(u)=1$. Thirty-eight of the 48 registered channel--epoch records passed the padding, completeness, and finite-waveform checks; characteristic-function reliability was then applied at each Fourier argument. Nine epochs contributed to the Livingston primary, giving $r_{\rm L1}(2)=0.5556$ with hierarchical-bootstrap 95\% interval $[0.5158,0.6425]$ (Fig.~\ref{fig:SO3}). The paired Livingston-minus-Hanford contrast used seven epochs and its interval contained zero. The slow characteristic-function shape therefore replicates across disjoint intervals at both sites, and the paired site contrast shows no Livingston-specific excess. Exploratory tail ablation moves both sites close to $r(2)=1$, indicating that rare one-second excursions carry most of the effect. These uncalibrated records establish a distribution-shape result.

\FloatBarrier

\begin{figure}[ht]
\centering
\subfloat[]{\includegraphics{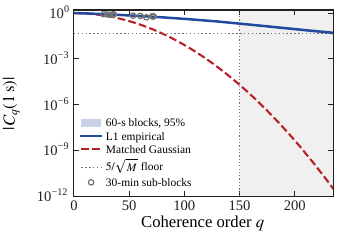}}\hspace{3pt}%
\subfloat[]{\includegraphics{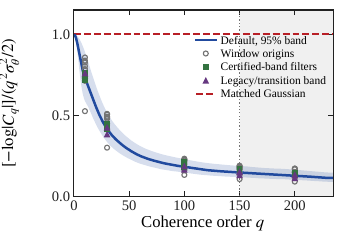}}\hspace{3pt}%
\subfloat[]{\includegraphics{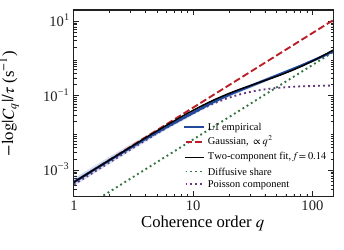}}
\caption{\label{fig:Sligodirect} Phase-equivalent audit of the GW170814 Livingston record for the assumed coupling $\gamma/(2\pi)=7~\mathrm{Hz\,nT^{-1}}$. (a)~$|C_q(1\,\mathrm{s})|$ with its 60-s-block-bootstrap band, the matched Gaussian, the $5/\sqrt M$ scale, and the six 30-min sub-block estimates (open circles); the reported range ends at $q=150$. (b)~Exponent normalized by the Gaussian exponent, with the window-origin (open circles), certified-band filter (squares), and legacy/transition-band (triangles) variants. (c)~Finite-window exponent $-\log|\widehat C_q|/\tau$ with its 60-s-block-bootstrap band, the matched Gaussian $q^2\sigma_\theta^2/2$, and the two-component fit of Eq.~\eqref{eq:Sdecomp} with its diffusive and Poisson parts.}
\end{figure}

A separate GW170814 auxiliary-channel release~\cite{S_GWOSC}, calibrated in the frequency band used below, permits a phase-equivalent diagnostic after choosing a magnetic coupling. We assume the coupling $\gamma/(2\pi)=7~\mathrm{Hz\,nT^{-1}}$. For nonoverlapping windows of duration $\tau$ we form
\begin{equation}
\theta_j(\tau)=\gamma\int_{j\tau}^{(j+1)\tau}B_{\rm filt}(t)\,\dd t,\qquad \widehat C_q(\tau)=\frac{1}{M}\sum_{j=1}^{M}e^{-iq\theta_j(\tau)}.
\label{eq:Sligodirect}
\end{equation}
For the prespecified transient-rich Livingston end-X floor channel at $\tau=1$~s, $|\widehat C_{100}|=0.4184$ $[0.3957,0.4401]$ and $|\widehat C_{150}|=0.2046$ $[0.1871,0.2218]$, compared with matched-variance Gaussian values $0.00820$ and $2.02\times10^{-5}$. Window-origin and certified-band filter variants preserve the qualitative failure of the Gaussian extrapolation through $q=150$ (Fig.~\ref{fig:Sligodirect}). The quantity $-\log|\widehat C_q(\tau)|/\tau$ is a finite-window exponent. Thresholded-event catalogs and their Poisson surrogates are not used because the inferred marks, void probabilities, and decompositions are detector- and preprocessing-dependent. Figure~\ref{fig:Sligodirect}(c) shows the finite-window exponent against the order. Its local logarithmic slope falls from $1.73$ $[1.64,1.81]$ at $q=10$ to $1.24$ $[1.15,1.33]$ at $q=50$ and rises to $1.57$ $[1.45,1.68]$ at $q=150$, between the Gaussian value $2$ and the saturated value $0$. The two-component form
\begin{equation}
\gamma_q=\frac{\sigma_\theta^2}{2}\,q^2\left[f+(1-f)\,h(qa)\right],\qquad h(x)=\frac{1}{1+x^2},
\label{eq:Sdecomp}
\end{equation}
which combines a diffusive share $f$ with a compound-Poisson component of Laplace-distributed kicks of scale $a$ and reproduces the exact small-$q$ limit, fits the exponent for $q\le150$ with a root-mean-square residual of $0.055$ in $\log\gamma_q$. It gives $f=0.14$ $[0.11,0.17]$, $a=0.046$~rad, and an implied event rate of $0.19$~s$^{-1}$ $[0.12,0.29]$; Gaussian-distributed kicks give $f=0.15$ with residual $0.085$. In the language of Sec.~S6, $f$ is the diffusive share of this record at the assumed coupling. The Gaussian part of the increments, which alone would set the asymptotic floor of any parallel protocol, carries $14\%$ of the one-second variance. The decomposition describes a finite-window exponent and depends on the coupling and the filter band.

\FloatBarrier

\begin{figure}[ht]
\centering
\subfloat[]{\includegraphics{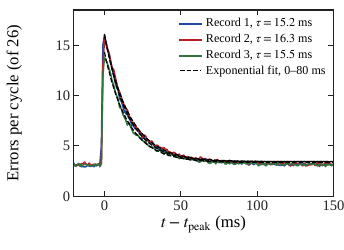}}\hspace{3pt}%
\subfloat[]{\includegraphics{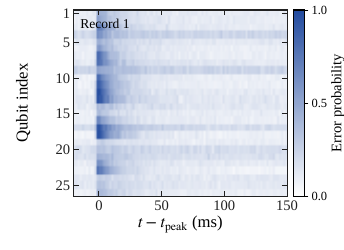}}\hspace{3pt}%
\subfloat[]{\includegraphics{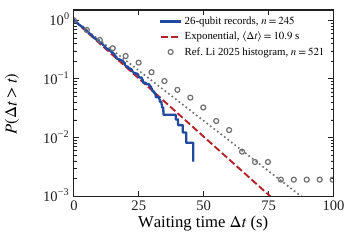}}
\caption{\label{fig:Smcewen} Quasiparticle bursts in the released 26-qubit records. (a)~Total errors per cycle around the selected peak of each curated record ($3\,\mu$s cycles, 0.999-ms bins), with exponential recovery fits over the first 80~ms. (b)~Per-qubit error probability in the first curated record; all 26 qubits participate. (c)~Survival function of the waiting times between bursts in the 60-s records against the exponential law of the same mean, and the binned waiting times between charge-parity-jump bursts released with Ref.~\cite{S_Li2025}, with their exponential fit (dotted).}
\end{figure}

\emph{Common events in a superconducting array.}
Radiation impacts in a superconducting processor create quasiparticle bursts that raise the error rate of every qubit within a chip-scale region for about ten milliseconds~\cite{S_McEwen}. Muon detectors operated synchronously with an array identify the cosmic-ray subset of these events~\cite{S_Harrington2025,S_Li2025}, so the events are heralded by external detectors as well as by their correlated-error signature. We reanalyzed the released 26-qubit records of Ref.~\cite{S_McEwen}, three curated records with 3-$\mu$s cycles that contain one strong burst each and 91 records of 60~s with 100-$\mu$s cycles. In the curated records, 0.999-ms bins show the burst peaks rising from off-burst levels near $3.1$ errors per cycle to $13.8$--$16.0$, with fitted recovery times of $15.2$--$16.3$~ms, and sliding windows over raw cycles show participation of all 26 qubits within 6--9~$\mu$s [Figs.~\ref{fig:Smcewen}(a) and \ref{fig:Smcewen}(b)]. Leave-one-qubit-out peak selection returns the same peak bin in all 78 tests, and the first principal component of the binned trajectories explains $0.72$--$0.94$ of the variance, above independently shifted nulls. In the 60-s records, a threshold detector (Sec.~S13) finds 332 bursts in 5460~s, a mean rate of 0.061~s$^{-1}$. Their footprints range from a few qubits to the whole array, with a median of 18 qubits, and 50 events involve all 26. The 245 waiting times within the records are consistent with the exponential law of the same mean [Kolmogorov--Smirnov $p=$0.81; Fig.~\ref{fig:Smcewen}(c)], and counts in 10-s windows have Fano factor 0.96, the Poisson value being 1. The 521 waiting times between charge-parity-jump bursts released with Ref.~\cite{S_Li2025} for a different processor are also exponential, with mean $14$~s [Fig.~\ref{fig:Smcewen}(c)]. These records establish the two properties that the model assigns to common events, participation of the whole array within microseconds for the strongest events and Poisson arrival, and they give the event duration of about 15~ms below which an event cannot be treated as an instantaneous kick. Rates of order $0.1$~s$^{-1}$ are five orders of magnitude below the single-qubit dephasing rates of present transmons, so for metrology these events matter through their heralded structure and not through the window of Sec.~S6; discarding heralded cycles converts them into erasures in the sense of Sec.~S5.

\FloatBarrier

\section{Numerical and data-analysis methods}
\label{sec:Smethods}
\emph{Numerical propagation and optimization.} In the Dicke basis, common dephasing multiplies each element of the initial density matrix by the factor in Eq.~\eqref{eq:Sreduction}. The QFI is evaluated from the spectral decomposition $\rho=\sum_j\lambda_j|j\rangle\langle j|$ as
\begin{equation}
 F_Q=2\sum_{\lambda_j+\lambda_k>0}
 \frac{|\langle j|\partial_\omega\rho|k\rangle|^2}{\lambda_j+\lambda_k},
 \qquad \partial_\omega\rho=-it[J_z,\rho],
 \label{eq:SnumericalQFI}
\end{equation}
in two independent implementations, with the analytic GHZ expression as a check. With amplitude damping, the full $2^N$ density matrix is built from the local Kraus operators of the damping channel followed by the commuting dephasing channel; this verifies Eq.~\eqref{eq:ST1QFI} for $N=1,2,4,6$, including population leakage. The probe optimizations behind main-text Fig.~2(a) and the finite-$N$ search are specified in Sec.~S5. The population equations~\eqref{eq:Scascade} of Fig.~\ref{fig:Ssuperradiance} are integrated with a stiff solver from the fully excited Dicke state and compared with a full-Liouvillian propagation that averages the pulse area by quadrature. The L\'evy integrals of Sec.~S4 are evaluated by quadrature with separate treatment of the origin singularity and the oscillatory tail.

\emph{Branch schematic and simulated parity measurements.} Main-text Fig.~3(a) is a schematic of the conditional decomposition of Secs.~S3 and S5: the no-kick branch carries the weight $p_0=e^{-\Gamma t}$, and the kicked branches are drawn as a uniform pedestal whose residual coherence $e^{-\Gamma t}[e^{\Gamma t\varphi(N)}-1]$ vanishes as $\mu N\to\infty$. No numerical data enter this panel. For Fig.~3(b), each of the 200 shots per point draws a Poisson number of Laplace kicks, sums them to a phase $\Theta$, and samples parity with probability $[1+\cos(q\Theta)]/2$; the contrast is the sample mean of the $\pm1$ outcomes, and the error bars are its standard error.

\emph{Common-noise records.} For each platform, increments over consecutive windows of fixed duration are centered and standardized by their sample standard deviation. The empirical characteristic function $\widehat C(u)$ and the exponent ratio of Eq.~\eqref{eq:Sshape} are evaluated where $|\widehat C(u)|$ exceeds its statistical reliability floor. Confidence intervals come from block bootstrap over contiguous segments long compared with the window, each replicate recomputing the standardization and the characteristic functions. Tail ablation removes the largest absolute increments and restandardizes the remainder; it describes the conditional central distribution and is kept separate from the full-record estimates.

For the fiber-delivered laser record~\cite{S_Schioppo2022}, the released one-second values of the fractional transfer beat are used as provided, including the authors' flags of invalid seconds; windows are formed only from consecutive valid seconds, and excursions are values beyond $4\sigma$, grouped into clusters when closely spaced. The phase change per second follows from the fractional value at the 194.4-THz carrier. The Allan deviation of the record is consistent with the instability reported in Ref.~\cite{S_Schioppo2022}.

For the O3 magnetometer records, the intervals, channels, and analysis rules were fixed before the analysis waveforms were accessed. The uncalibrated records are anti-alias filtered and downsampled, the low-frequency band and the mains harmonics are removed by Fourier filtering, and the result is integrated over one-second windows; only complete, uniformly sampled records are used, without gap filling. The Livingston primary and the paired site contrast are medians over the eligible epochs, with hierarchical bootstrap over epochs and blocks. Sensitivity checks vary the window origin, the filter cutoffs, and the block length.

The GW170814 diagnostic uses the calibrated Livingston end-station magnetometer with the same filtering, the assumed coupling of Sec.~S12, which converts field integrals into the phase increments of Eq.~\eqref{eq:Sligodirect}, and a Gaussian reference of matched variance. The reported order range is limited by the reliability floor of $|\widehat C_q|$ and by the filter checks, and Hanford channels serve as controls. For Fig.~\ref{fig:Sligodirect}(c), local exponents are least-squares slopes of $\log\gamma_q$ over a narrow band of orders around $q$, and Eq.~\eqref{eq:Sdecomp} is fitted by least squares in $\log\gamma_q$, both with block-bootstrap intervals; the implied event rate is $(1-f)\sigma_\theta^2/(2a^2)$.

For the superconducting-array records, errors are binned in time and summed over qubits. In the curated records the burst peak maximizes the summed error rate, the baseline is the off-burst median, and the recovery is fitted by an exponential; leave-one-qubit-out peak selection and principal-component analysis against circularly shifted null trajectories test the collective structure. In the 60-s records a threshold detector on the binned error count marks burst starts after a quiet interval, participation is assigned from each qubit's excess error probability after the peak, and waiting times between successive starts within a record are compared with the exponential law of the same mean by a Kolmogorov--Smirnov test; censoring at the record ends makes their mean smaller than the inverse rate. The histogram of Ref.~\cite{S_Li2025} is taken from the source data of its Fig.~3e without reanalysis.

Data sources are cited in Sec.~S12; the analysis code and derived records are available from the authors.

\bibliography{SM_References}